\documentclass[%
 reprint,
 amsmath,amssymb,
 aps,
]{revtex4-2}

\usepackage{graphicx}
\usepackage{dcolumn}
\usepackage{bm}
\usepackage{color}

\begin{document}

\preprint{APS/123-QED}

\title{Unravel the rotational and translational behavior of a single squirmer in flexible polymer solutions at different Reynolds numbers}
\author{Yuan Zhou}
\affiliation{2020 X-Lab, Shanghai Institute of Microsystem and Information Technology, Chinese Academy of Sciences, Shanghai 200050, China}
\affiliation{College of Materials Science and Opto-Electronic Technology, University of Chinese Academy of Sciences, Beijing, 100049, China}
\author{Kai Qi}
\email{kqi@mail.sim.ac.cn}
\affiliation{2020 X-Lab, Shanghai Institute of Microsystem and Information Technology, Chinese Academy of Sciences, Shanghai 200050, China}
\affiliation{College of Materials Science and Opto-Electronic Technology, University of Chinese Academy of Sciences, Beijing, 100049, China}
\affiliation{Centre Europ\'een de Calcul Atomique et Mol\'eculaire (CECAM), \'Ecole Polytechnique F\'ed\'erale de Lasuanne (EPFL), Batochime, Avenue Forel 2, 1015 Lausanne, Switzerland}
\author{Marco De Corato}
\affiliation{Aragon Institute of Engineering Research (I3A),  University of Zaragoza, Zaragoza, Spain}
\author{Kevin Stratford}
\affiliation{Edinburgh Parallel Computing Centre, The University of Edinburgh, The King’s Buildings, Edinburgh EH9 3JZ, Scotland}
\author{Ignacio Pagonabarraga}
\email{ipagonabarraga@ub.edu}
\affiliation{Departament de F\'isica de la Mat\`eria Condensada, Universitat de Barcelona, C. Mart\'i Franqu\`es 1, 08028 Barcelona, Spain}
\affiliation{Universitat de Barcelona Institute of Complex Systems (UBICS), Universitat de Barcelona, 08028 Barcelona, Spain}
\affiliation{Centre Europ\'een de Calcul Atomique et Mol\'eculaire (CECAM), \'Ecole Polytechnique F\'ed\'erale de Lasuanne (EPFL), Batochime, Avenue Forel 2, 1015 Lausanne, Switzerland}

\begin{abstract}
Microorganisms, such as bacteria and algae, thrive in complex environments and their behavior in fluids holds significant importance for various medical and industrial applications. By conducting Lattice Boltzmann (LB) simulations, the transport and rotational properties of a generic squirmer are investigated in solutions embedded with flexible polymer chains at different Reynolds numbers. The interplay of activity and heterogeneously distributed polymers have profound influences on these properties. Remarkable enhancements of up to three orders of magnitude in the rotational motion, along with apparent decays in self-propelling velocities, are observed for squirmers with non-zero active stresses. These extraordinary phenomena stem from the squirmer-polymer mechanical and hydrodynamic interactions. Specifically, polymer wrapping occurs in front of a pusher, while numerous polymers are absorbed in the rear of a puller. Both mechanisms enhance the rotational motion and simultaneously impede translations through forces and torques arising from direct contacts or asymmetric local flows induced by polymers. The source dipole flow fields generated by a neutral swimmer rapidly advect polymers to the rear, leaving no apparent impacts on its rotational and transport properties. The influences of Reynolds number {\rm Re} ({\rm Re=0.8} and 0.04) and squirmer-polymer boundary conditions (no-slip and repulsive) on the dynamics are addressed. In short, the no-slip boundary condition results in more profound effects on both rotational and translational properties at ${\rm Re} = 0.8$. However, at ${\rm Re} = 0.04$, the disparity between the two boundary conditions diminishes due to the heightened fluid viscous drag, which impedes direct contacts between squirmers and polymers. Our results reveal the relevance of system heterogeneity and highlight the essential role of squirmer-polymer mechanical and hydrodynamic interactions in shaping the behavior of swimmers in viscoelastic fluids. These findings offer valuable insights for the design and control of artificial microrobots operating in complex environments.
\end{abstract}

\maketitle


\section{Introduction}

In contrast to simple viscous fluids, the natural habitats of microorganisms and the operational environments of artificial microswimmers are generally complex. These environments are typically viscoelastic and span various Reynolds number regimes. 

As a dimensionless quantity, the Reynolds number (Re) predicts fluid flow patterns in different circumstances by measuring the ratio between inertia and viscous forces \cite{reynolds1883xxix,rott1990note}. In general, microswimmers typically reside in the regime of extremely low Reynolds numbers (${\rm Re} \ll 1$), displaying a rich array of behaviors. For instance, \emph{Bacillus subtilis} and \emph{Chlamydomonas} swim at Reynolds numbers of approximately $10^{-4}$ \cite{wolgemuth2008collective} and $10^{-3}$ \cite{bennett2015steering}, respectively, while \emph{Paramecium} swims at ${\rm Re} \approx 0.1$ \cite{brette2021integrative}. In the low Reynolds number range ($\rm Re = 10^{-5}$ to $10^{-3}$), densely suspended bacteria display collective behaviors \cite{li2016hydrodynamic} such as large-scale flow patterns and vortices \cite{dombrowski2004self}, locally synchronized movements \cite{sokolov2007concentration}, uneven spatial distribution of swimmers, and increased diffusion and mixing \cite{wu2000particle}. William et al. first reported a microscale, biohybrid swimmer propelled by contractile cells, operating at a low ${\rm Re}\approx10^{-2}$. By using the standard lithographic techniques and mass-scale cell plating, these swimmers are amenable to batch fabrication \cite{william2014self-propelled}. At a low Reynolds number $(\rm Re=2.5 \times 10^{-4})$, the swimming of \emph{Chlamydomonas reinhardtii} cells experience higher viscous friction during forward motion than backward motion, breaking time-reversal symmetry and ensuring propulsion \cite{garcia2011random}. A study on low Reynolds number propulsion ($1.4 \times 10^{-4} \sim 3 \times 10^{-3}$) of a scallop indicates that differences in the opening and closing rates result in varying shear rates and, consequently, different viscosities in non-Newtonian fluids \cite{qiu2014swimming}. In addition, the average velocity of the micro-scallop in the shear thickening fluid is faster than in a shear-thinning fluid \cite{qiu2014swimming}. Another study on a flagellated swimmer $(\rm Re<0.3)$ in unbounded space driven by Quincke rotation reveals that the Quincke swimmer exhibits three forms of motion—roll, pitch, and self-oscillatory rotation—by varying the electric field and the angle between the two filaments \cite{endao2021biflagellated}.

In contrast to the low Reynolds number regime where viscous forces are dominant, the influence of inertia on the behavior of microswimmers has gained increasing attention. Investigations at finite Reynolds numbers not only shed light on predator-prey interactions, sexual reproduction, and encounter rates of marine organisms but also aid in the design and fabrication of efficient artificial swimmers \cite{park2016phototactic,feldmann2021can}. Squirmers with different swimming schemes exhibit significant differences in locomotion across Reynolds number $(0.01 <\rm Re < 1000)$, i.e., pushers display a monotonic increase in swimming speed with raising Re, whereas pullers exhibit a non-monotonic behavior \cite{chisholm2016squirmer}. By using the squirmer model in the regime of ${\rm Re} \sim {\rm O}(0.1–100)$, Li revealed that inertial effects alter the contact time and dispersion dynamics of a pair of pusher swimmers, while triggering hydrodynamic attraction between two pullers \cite{li2016hydrodynamic}. A study on a self-propelled slender swimmer in a chaotic flow field with a flow Reynolds number up to 10 illustrated that pushers exhibit more efficient locomotion than pullers due to the different distribution of vorticity within the wake \cite{cavaiola2021self}. Suspensions of slender pusher and puller swimmers present nontrivial flow motions as the Reynolds number $(1 <\rm Re < 50)$  changes along with complex swimmer dynamics \cite{cavaiola2022swarm}. The pairwise hydrodynamic interactions for simple two-dimensional dimer model swimmers \cite{dombrowski2020kinematics,dombrowski2019transition,derr2022reciprocal} over a range of finite Re were investigated numerically. Based on the Reynolds number $(0.1 <\rm Re < 40)$ and initial positions, two swimmers can either repel and move away from each other or form one of four stable pairs: in-line and in tandem \cite{dombrowski2022pairwise}. An experimental and numerical study on self-assembled ferromagnetic spinners at ${\rm Re}\approx30$ showed that spinner suspensions induce vigorous vortical flows at the interface, exhibiting properties of well-developed 2D hydrodynamic turbulence \cite{kokot2017active}.

The natural habitats of microorganisms and spermatozoa predominantly consist of polymeric environments
\cite{li2021microswimming}. For instance, sperm exhibit motility in the cervix and along the fallopian tubes \cite{katz1978movement,katz1980flagellar,katz1981movement,suarez1992hyperactivation,suarez2006sperm, fauci2006biofluidmechanics,hyakutake2015effect}, while bacteria swarms within biofilms composed of extracellular polymeric substances (EPS) \cite{o2000biofilm,donlan2002biofilms,costerton1987bacterial,costerton1995microbial,wilking2011biofilms, yazdi2012bacterial,bar2012revised,karimi2015interplay}. Helicobacter pylori, a bacterium associated with ulcers, adapts to the acidic conditions of the stomach by altering the rheological properties of the mucus lining
\cite{montecucco2001living,celli2009helicobacter}. Additionally, synthetic micromotors employed in drug delivery \cite{luo2018micro}, microsurgery \cite{ullrich2013mobility}, and disease detection \cite{wang2020review} operate within complex environments. As a consequence, comprehending the locomotion of swimmers in non-Newtonian fluids is imperative for potential biomedical and industrial applications. Intuitively, the existence of macromolecules will hinder the translational motion of swimmers due to increased viscosity \cite{shen2011undulatory,zhu2012self,qin2015flagellar,datt2017active}. However, abundant variations in swimming speed have been reported \cite{patteson2015running,berg1979movement,teran2010viscoelastic,espinosa2013fluid, martinez2014flagellated,jung2010caenorhabditis,leshansky2009enhanced,gomez2017helical,magariyama2002mathematical, zottl2019enhanced}. By using finite element methods, Zhu et al. \cite{zhu2012self,zhu2011locomotion} revealed that both translational motion and power consumption of squirmers are reduced in a polymeric fluid. In the case of two-dimensional undulatory kicker swimmers, the high polymer stress at the tail enhances the swimming velocity \cite{thomases2014mechanisms}. The rotating flagella of bacteria generate a depletion zone of long polymers, resulting in an apparent slip velocity between the fluid and the swimmer. Consequently, the swimming speed has been increased up to $60\%$ \cite{zottl2019enhanced}. Notably, recent experiments have reported significant enhancements in the rotational motion of active Brownian colloidal particles embedded in viscoelastic fluids \cite{gomez2016dynamics}. Subsequent studies demonstrated that the enhanced diffusions turn into persistent circular rotations above a critical Deborah number \cite{narinder2018memory}. Meanwhile, hydrodynamic simulations via multiparticle collision dynamics indicated that a decrease of adsorbed polymers by active motion and asymmetric squirmer-polymer encounters can result in large rotational enhancements for a neutral squirmer \cite{qi2020enhanced}. In addition, the interplay between geometrical confinement and fluid viscoelasticity gives rise to intriguing phenomena. Recent experiments unraveled that \emph{E. coli} immersed in a DNA solution and under spherical confinement can self-organize into a millimeter-scale rotating vortex. The giant vortex switches its global chirality periodically with tunable frequency \cite{liu2021viscoelastic}. Experiments involving active particles in viscoelastic fluids under various geometrical constraints, such as flat walls, spherical obstacles, and cylindrical cavities, revealed that confined viscoelastic fluids can induce an effective repulsion on particles when approaching a rigid surface. The repulsion strength has a dependence on the incident angle, surface curvature, and particle activity \cite{narinder2019active}. In the case of swimmer suspensions, viscoelasticity exerts profound influences on collective behavior \cite{bozorgi2011effect,bozorgi2013role,bozorgi2014effects,li2016collective}. For example, Bovine spermatozoa exhibit disordered individual swimming in Newtonian fluids, but cell-cell alignments and dynamic cluster formations in viscoelastic fluids \cite{tung2017fluid}. Additionally, simulations of multiple rodlike microswimmers in 2D continuum viscoelastic fluids demonstrated that the cluster aggregation for pushers is evidently enhanced by viscoelasticity, while the effect on pullers is subtle \cite{li2016collective}.

However, the interplay of viscoelasticity and Reynolds number on the dynamics of microswimmers has not been thoroughly investigated. In this paper, we aim at revealing the translational and rotational mechanisms of generic spherical squirmers immersed in viscoelastic fluids at different Reynolds numbers by employing the Lattice-Boltzmann method. Viscoelasticity is incorporated by taking linear flexible polymers explicitly into account. Both the no-slip and repulsive boundary conditions between the squirmers and polymers are examined, with a specific focus on the influence of the fluid Reynolds number. In the low ${\rm Re}=0.04$ regime, a significant enhancement of rotation diffusion, exceeding 3 orders of magnitude, for the puller has been observed, whereas the translational motion of all types of squirmers is reduced due to the increased viscous drag and direct hindrance caused by polymers. The micromechanism underlying the rotational enhancement depends on the active stress. To be more specific, a polymer wrapping is formed in front of a pusher and numerous polymers are asymmetrically absorbed in the rear of a puller. These mechanisms result in either asymmetric collisions between polymers and squirmers or induce heterogeneous flows, which implicitly increase the rotational motion via the fluid-swimmer no-slip boundary condition. In contrast, polymers are rapidly advected by the flow in the vicinity of a neutral swimmer, resulting in no evident impacts on its rotational property. At ${\rm Re}=0.04$, the differences between no-slip and repulsive boundary conditions in the transport and rotational diffusion of squirmers are small. Because the viscous drag of fluid decelerates the advection of polymers to the vicinity of squirmers, leading to fewer collisions. On the contrary, at ${\rm Re}=0.8$, the rotational enhancement under no-slip boundary condition is up to 4 times stronger compared to the repulsive case, indicating the relevance of fluid inertia in facilitating the advection of polymers. The influences of polymer concentration, length, and polymer-squirmer aspect ratio on the rotational and translational motion of swimmers are also investigated. Our findings highlight the significance of system heterogeneity resulting from the presence of polymers in determining the behavior of squirmers in viscoelastic fluids.

This paper is structured as follows: Sec.~II describes the numerical strategy employed to model the polymer suspensions and squirmers in terms of a state-of-the-art Lattice Boltzmann simulation method. Since the mechanisms associated with the transport and rotational properties of squirmers are more prominent at ${\rm Re} = 0.8$, the influence of fluid inertia is first discussed in Sec.~III. Afterwards, the relevance of fluid viscous drag on squirmers’ motion at ${\rm Re} = 0.04$ is addressed in Sec.~V and IV. Finally, Sec.~VI draws conclusions.

\section{Model and simulation methods}

\begin{figure}
  \centering
  \includegraphics[width=0.7\columnwidth]{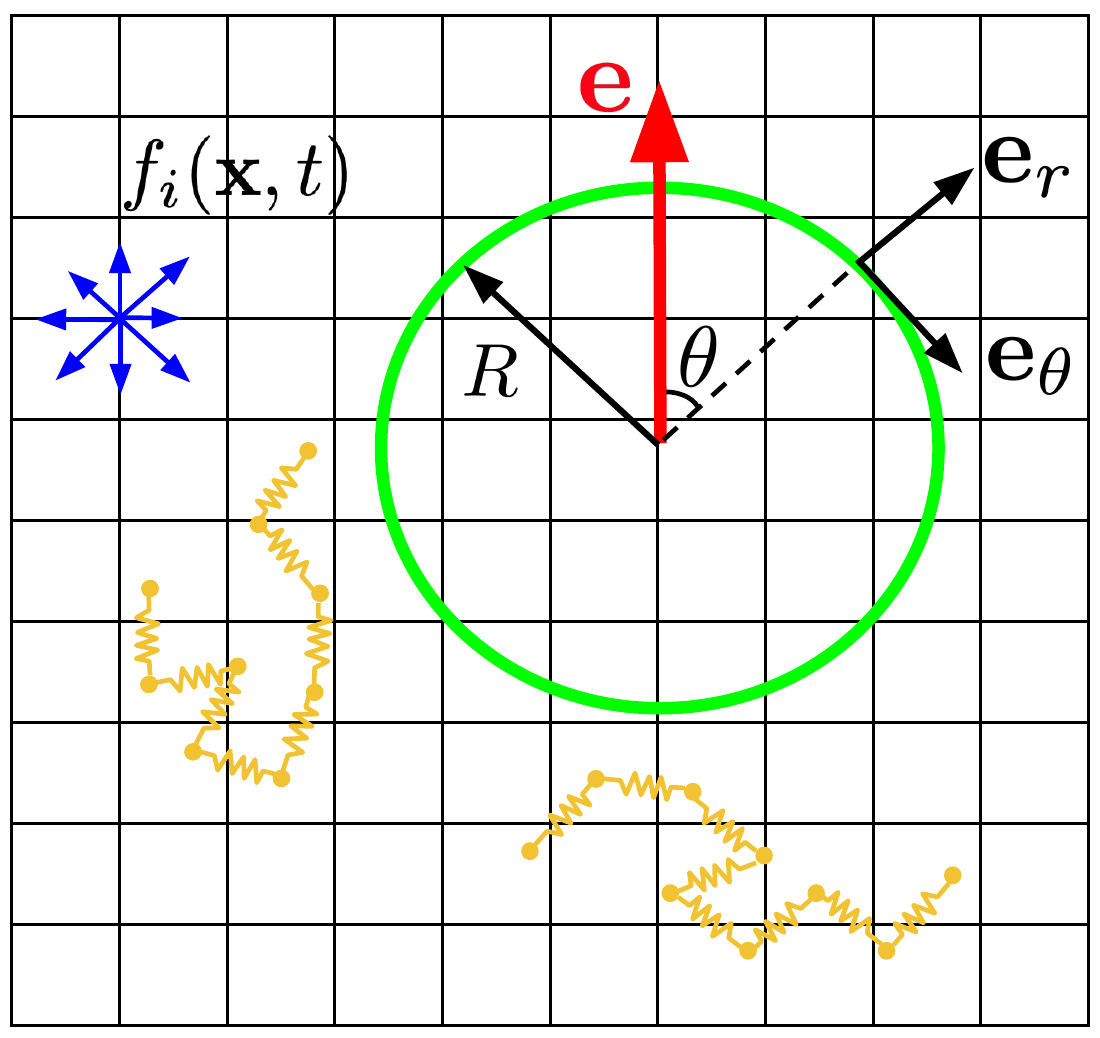}
    \caption{
        Sketch of a spherical squirmer with radius $R$ immersed in a viscoelastic fluid.
        The squirmer exhibits self-propulsion along the polar direction ${\bf e}$, with 
        ${\bf e}_r$ and ${\bf e}_{\theta}$ representing the radial and tangential unit vectors, 
        respectively. The fluid one-particle distribution function $f_i({\bf x},t)$ can only propagate
        along the lattice. The viscoelasticity is introduced by embedding flexible polymers (yellow
        bead-spring chains) into the fluid.
    }
  \label{sketch}
\end{figure}

The mesoscopic fluid environment is achieved by performing Lattice-Boltzmann (LB) simulations \cite{dunweg2009lattice,nash2008singular,pozrikidis1992boundary,peskin2002immersed}, where the D3Q19 lattice scheme \cite{stratford2008parallel} is employed. Unlike the fully resolved hard-sphere models with stick boundary conditions, which have been successfully implemented in colloid systems \cite{ladd1994numerical,ladd1994numerical311,nguyen2002lubrication}, flexible polymers are simulated by a bead-spring model in this work (Fig.~\ref{sketch}, supplemental materials). Monomers are modeled as point particles acting as force monopoles \cite{nash2008singular}, known as Stokeslets, exerting forces on the fluid. In contrast to previous works \cite{ahlrichs1999simulation,berk2005lattice}, monomer's inertia is also not considered. Because it does not play a significant role in determining the dynamics, especially the rotational motion, of swimmers in these complex systems. However, it is important to note that the inertia of the fluid is retained in simulations. As shown in Fig.~\ref{sketch}, a spherical squirmer \cite{lighthill1952squirming,blake1971spherical} is employed to model the microswimmer in this work. The swimming speed  $U_0=2/3B_1$ is determined by the first surface mode strength $B_1$ of the squirmer. The self-propulsion mechanism is governed by the active stress $\beta$, i.e., $\beta>0$ corresponds to a puller, $\beta=0$ to a neutral swimmer, and $\beta<0$ to a pusher \cite{qi2020enhanced,alarcon2013spontaneous,kuron2019lattice}. To prevent fluid from penetrating the squirmer, the no-slip boundary condition is enforced \cite{nguyen2002lubrication}. We have investigated the influences of two types of polymer-squirmer boundary conditions on the dynamics of the squirmer. Reminiscent of the fluid-squirmer coupling, a no-slip boundary condition between a squirmer and a monomer is first implemented. For comparison, the effect of the repulsive boundary condition on the behavior of the squirmer is also examined. Even though no net torque is generated during this process, the presence of polymers near a squirmer can induce asymmetric flows, thereby implicitly influencing its translational and rotational properties. In our work, all the quantities are expressed in lattice units. We set the characteristic distance, time, and mass as $a=\tau=m=1$. Details of the simulation method and models can be found in the supplemental materials.

In the following sections, the rotational and translational properties of various squirmers immersed in polymer solutions are investigated. The influence of fluid inertia and viscous forces is examined with the Reynolds numbers ${\rm Re}=2R\rho U_0/\eta=0.8, 0.04$, where two viscosities $\eta=0.025,0.5 m/a\tau$ are used. $R=3a$ is the radius of the squirmer and the fluid density $\rho=1 m/a^3$. To characterize the viscoelasticity of the fluids, the corresponding Deborah numbers ${\rm De}=U_0 \tau_p/2R=2789, 20944$ are estimated at temperature $k_BT= 10^{-4} ma^2/\tau^2$. Here the longest polymer relaxation times $\tau_p \approx 5\times 10^6, 3.8\times 10^7 \tau$ are measured by calculating the end-to-end vector correlation functions, which present an exponential decay \cite{huang2010semidilute}. In general, the fluid exhibits an elastic property if ${\rm De} \gg 1$, whereas if ${\rm De} \ll 1$, viscous forces dominate. However, the very large Deborah numbers in our work do not necessarily indicate strong elasticity within the fluids. In fact, the relaxation process of locally deformed polymers during self-propulsion is too slow for the elastic forces to significantly impact the squirmers. Therefore, the notable enhancements in rotational motion and the decrease in swimming velocities are primarily attributed to the mechanical and hydrodynamic interactions between squirmers and polymers. These points will be elucidated in the subsequent sections.


\section{Results}


\subsection{Transport and rotational properties at ${\rm Re}=0.8$}

\begin{figure}
  \centering
  \includegraphics[width=\columnwidth]{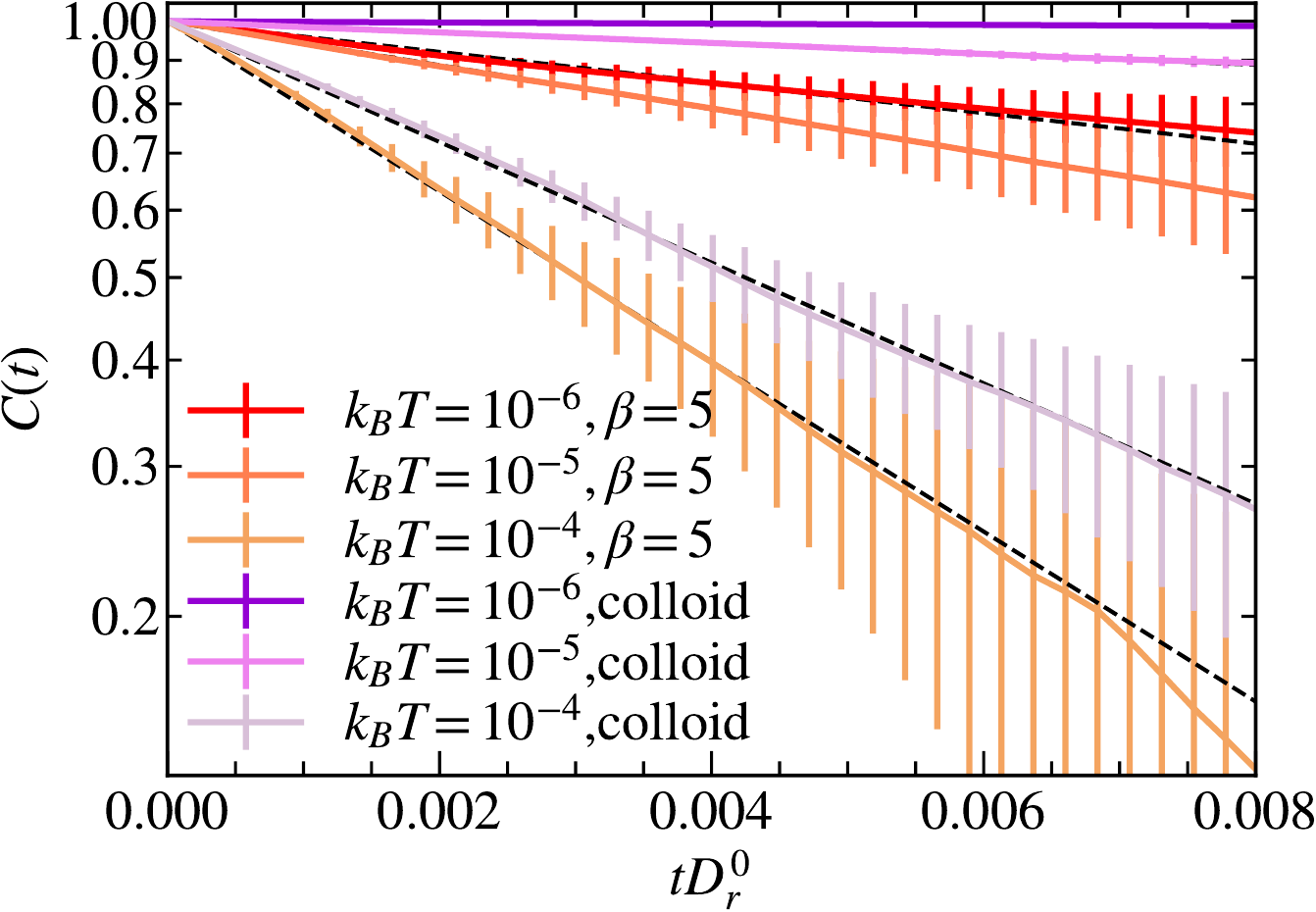}
    \caption{
        Orientation correlation functions for pullers ($\beta=5$) and colloids at different temperatures. 
        The Reynolds number is ${\rm Re}=0.8$, and a no-slip boundary condition is enforced between squirmers 
        and monomers. Solid lines are simulation results and dashed lines represent the fitted curves with 
        $D^{a,ns}_r=1.2, 1.8, 6.8 \times 10^{-6} \ \tau^{-1}$ for pullers and 
        $D^{p,ns}_r=4.7\times 10^{-8}, 4.3\times 10^{-7}, 4.8\times 10^{-6} \ \tau^{-1}$ for passive colloids
        at $k_BT/E_t=0.0015, 0.015, 0.15$, respectively. $D^0_r=5.9\times 10^{-8} \ \tau^{-1}$ 
        represents the theoretical estimation of the rotational diffusion coefficient of a colloid with 
        $R=3a$ at temperature $k_BT/E_t=0.0015$.
    }
  \label{Dr_beta5_rp}
\end{figure}

The rotational and translational behavior of squirmers with different swimming strategies in complex fluids is investigated at ${\rm Re}=0.8$. Viscoelasticity is achieved by embedding $N_p=96$ flexible polymers with a length of $N_m=240$ into the solutions. The corresponding monomer packing fraction is $\phi=N_mN_p/L^3=0.06 a^{-3}$. 

The rotational behavior of a squirmer is characterized by the orientation correlation function
\begin{equation}
    C(t)=\langle {\bf e}(t) \cdot {\bf e}(0) \rangle=e^{-2D_rt}, \label{orient_corr}
\end{equation}
where $D_r$ is the activity-dependent rotational diffusion coefficient.
Fig.~\ref{Dr_beta5_rp} shows the orientation correlation functions for pullers and colloids at various temperatures. We have performed $4 \sim 15$ independent runs with $2\times 10^6$ LB steps for each parameter set. The time average $\langle \cdots \rangle$ is taken when calculating $C(t)$,  and $D_r$ is determined by fitting the measured curves with Eq.~\ref{orient_corr}. Here, the no-slip boundary condition between squirmers and monomers is employed. As shown in Fig.~\ref{Dr_beta5_rp}, the rotational diffusion coefficients for passive colloids obey the Stokes–Einstein relation and are minimally affected by the presence of polymers. They exhibit a coherent change over two orders of magnitude as the temperature increases. However, activity significantly enhances the rotational diffusions for pullers at low temperatures. For example, $D^{a,ns}_r/D^{p,ns}_r \approx 26$ at $k_BT/E_t=0.0015$. Whereas, the thermal effect is dominant at high temperature $k_BT/E_t=0.15$ and the enhancement due to activity is subtle. Here $E_t=2/3\pi R^3 \rho U_0^2=6.3\times10^{-3} ma^2/\tau^2$ is the reference translational kinetic energy of the squirmer.

\begin{figure}
  \centering
  \includegraphics[width=\columnwidth]{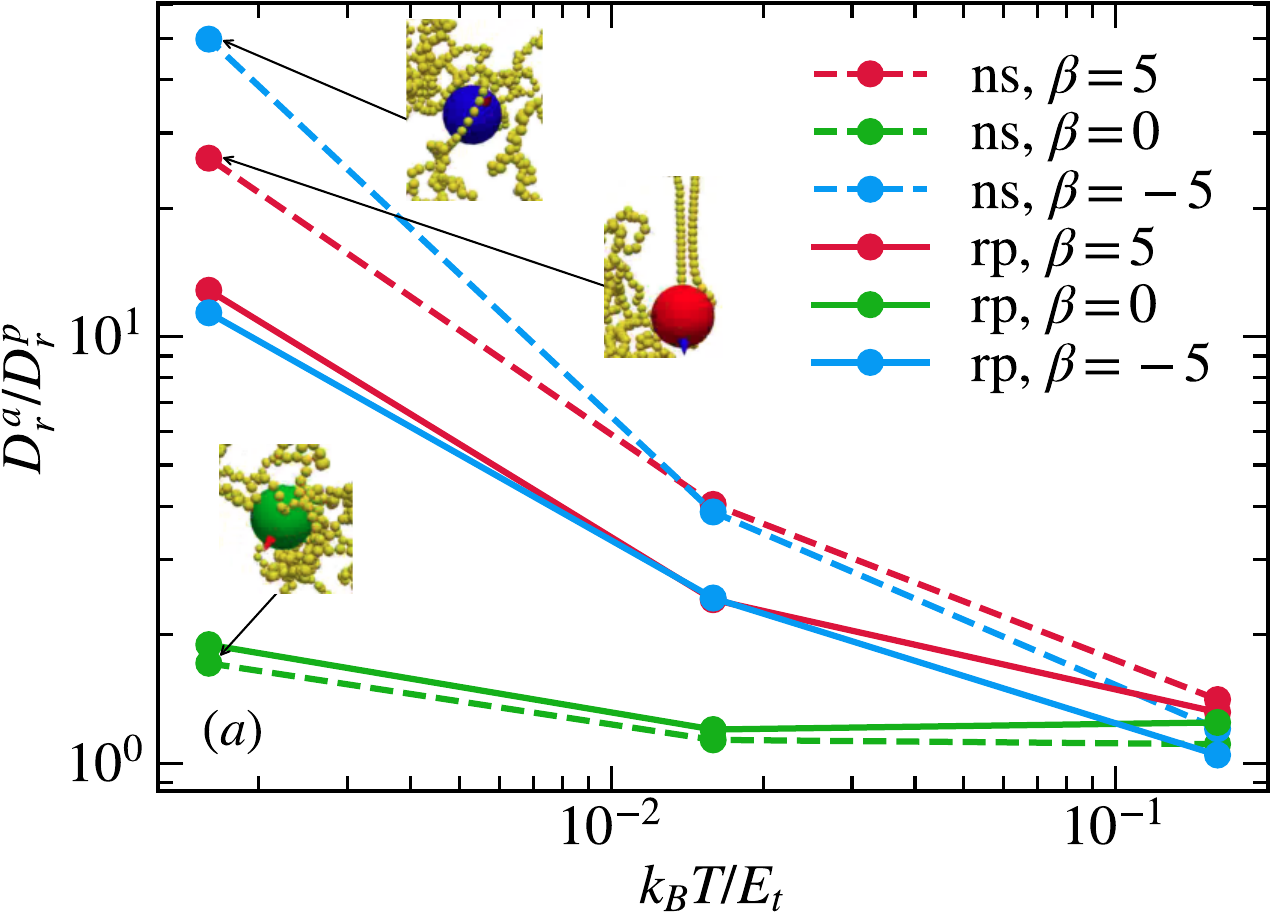}
  \includegraphics[width=\columnwidth]{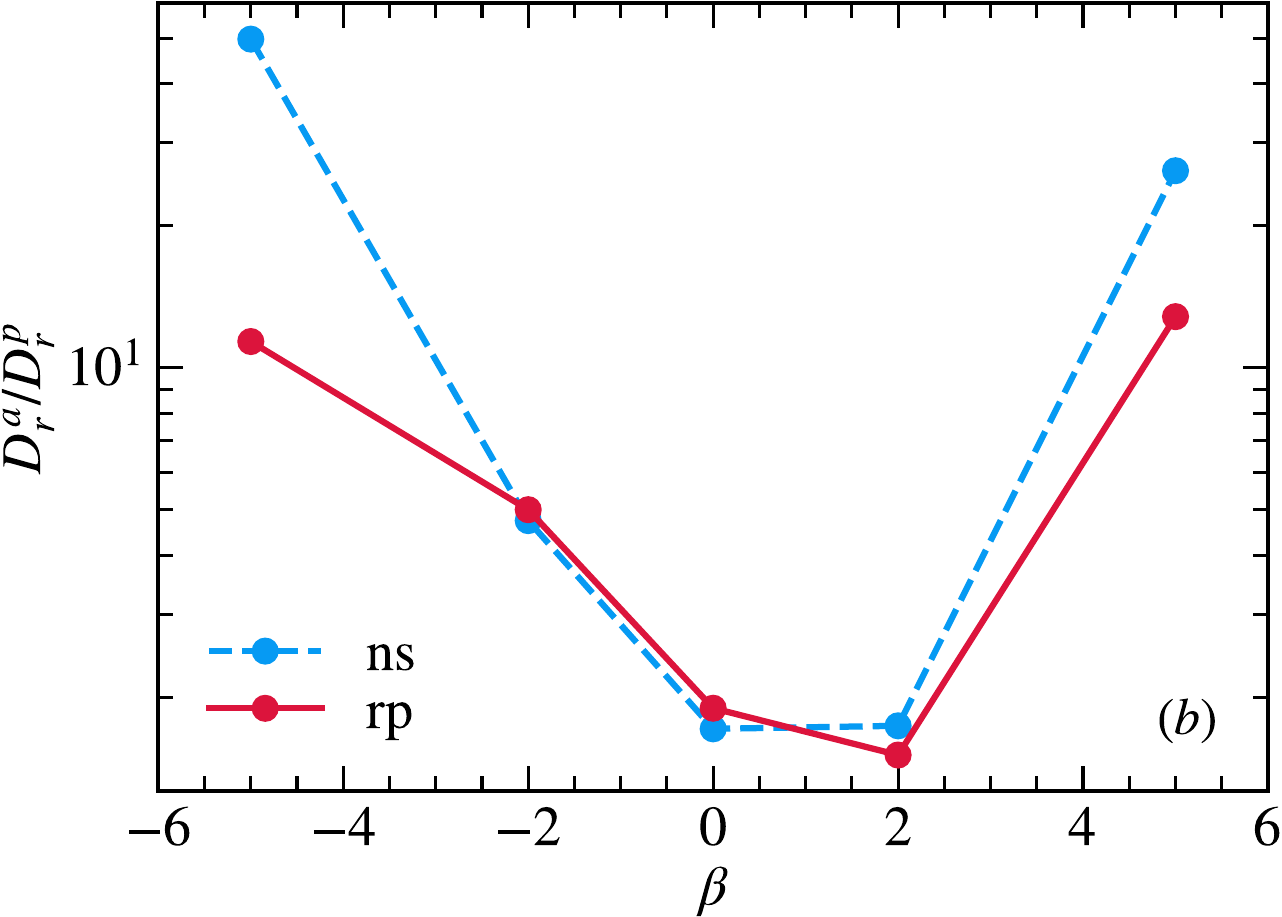}
    \caption{
        (a) Rotational diffusion enhancements for swimmers with the no-slip (ns) and the repulsive (rp)
        boundary conditions at ${\rm Re}=0.8$. Here, the rotational diffusion coefficients for passive colloids
        with the no-slip boundary condition are 
        $D^{p,ns}_r=4.7\times 10^{-8}, 4.3\times 10^{-7}, 4.8\times 10^{-6} \ \tau^{-1}$ at temperatures 
        $k_BT/E_t=0.0015, 0.015, 0.15$, respectively.
        Similarly, $D^{p,rp}_r=4.9\times 10^{-8}, 4.5\times 10^{-7}, 4.2\times 10^{-6} \ \tau^{-1}$ for
        the case with the repulsive boundary condition. Snapshots are the representative configurations 
        for squirmers with the no-slip boundary condition at $k_BT/E_t=0.15$. (b) Dependences of rotational 
        enhancements on the active stress at $k_BT/E_t=0.0015$.
    }
  \label{Dr_0p8}
\end{figure}

The rotational enhancements of different types of swimmers are summarized in Fig.~\ref{Dr_0p8}. At $k_BT/E_t=0.0015$, due to the interplay of activity and system heterogeneity caused by the presence of polymers \cite{qi2020enhanced}, profound enhancements near two orders of magnitude are observed for both pushers and pullers. However, the thermal fluctuation weakens this effect as temperature increases and becomes completely dominant at $k_BT/E_t=0.015$. On the contrary, neutral swimmers are insensitive to polymers. The rotational diffusion is only enhanced twice at $k_BT/E_t=0.15$. 

\begin{figure*}
  \centering
  \includegraphics[width=0.8\paperwidth]{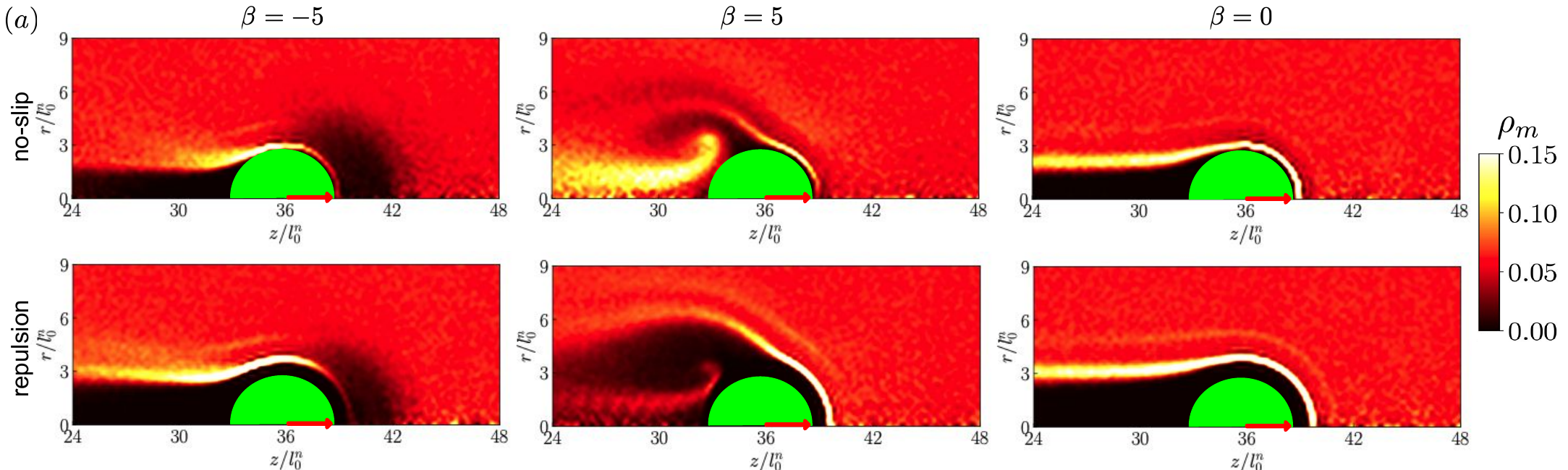}
  \includegraphics[width=0.8\paperwidth]{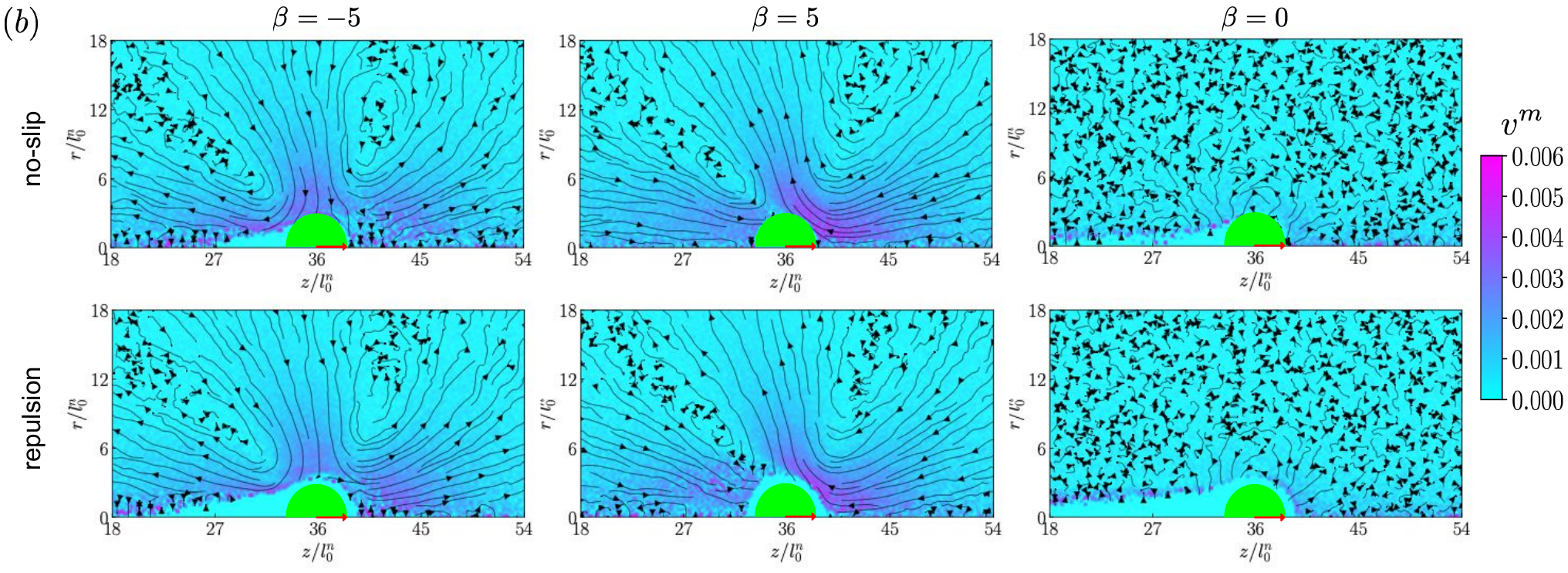}
    \caption{
        Cylindrical monomer (a) density and (b) velocity distributions of pushers ($\beta=-5$), 
        pullers ($\beta=5$), and neutral squirmers ($\beta=0$) at ${\rm Re}=0.8$. Here, the no-slip
        and repulsive boundary conditions are applied, respectively.}
  \label{cylin_0p8}
\end{figure*}

In Fig.~\ref{Dr_0p8}(a), the rotational diffusion coefficient of a pusher with the no-slip boundary condition is enhanced over 50 times at $k_BT/E_t=0.0015$. At ${\rm Re}=0.8$, where fluid inertia is relevant, the polymers are rapidly advected by the flow ($v^m \sim 0.0035 a/\tau$ in Fig.~\ref{cylin_0p8}(b)) to the vicinity of the squirmer, yielding a high monomer concentration on its surface. In particular, a distinct narrow ring with a monomer density of $\rho_m=0.09 a^{-3}$ can be observed in front of the no-slip pusher in Fig.~\ref{cylin_0p8}(a). This high density originates from the polymer wrapping effect during swimming. These polymers are stabilized on the squirmer's surface by the advective flow from the side. Consequently, a large amount of momentum and angular momentum exchanges occur, which greatly facilitate the rotational motion of the pusher (movie 1, Fig.~\ref{Dr_0p8}(a) sketch). This mechanism is the major contribution to its rotational enhancement. Despite the observation of a high monomer concentration at $\theta\approx \pi/2$, frequent collisions on both sides of the squirmer compensate for each other, resulting in zero net torque and thus no significant contribution to the rotational enhancement. However, when the repulsive boundary condition is applied, direct encounters between polymers and the squirmer are prohibited, weakening the wrapping effect (movie 2). Consequently, the monomer density in front of a pusher decreases to $\rho^m=0.04 a^{-3}$. But the heterogeneously distributed polymers can induce asymmetric flows, which indirectly mediate the interactions between the pusher and polymers through the fluid-squirmer no-slip boundary condition. Therefore, an enhancement of over 10 times on the rotational diffusion of a repulsive pusher is observed at $k_BT/E_t=0.15$. Naturally, due to the absence of direct encounters, this enhancement is 4.2 times weaker compared to the case with a no-slip boundary condition.

The rotational enhancement for a puller ($\beta=5$) is also profound. In the case with no-slip boundary condition, the enhancement is 26 times. As shown in movie 3, it is primarily attributed to the presence of heterogeneously distributed polymers absorbed in the rear of the puller. These polymers are advected to the vicinity of the puller by the backward suction flows (movie 3, Fig.~\ref{Dr_0p8}(a) sketch). Their presence induces asymmetric flows, which in turn enhance the squirmer’s rotation via the fluid-squirmer no-slip boundary condition. Additionally, collisions between the squirmers and polymers contribute additional asymmetrical forces and torques, further facilitating the rotational motion of the puller. In Fig.~\ref{cylin_0p8}(a), a widely distributed monomer density $\rho=0.15 a^{-3}$ is observed in the rear of the puller. Meanwhile, abundant polymers also appear at the front. However, they have a weak influence on the squirmer's rotation. Because on one hand, due to the high frequency of the polymer-squirmer collisions, these encounters are homogeneous. Thus, no net momentum or angular momentum is transferred to the puller. On the other hand, these polymers are rapidly advected to the side of the puller, exerting a weak influence on its rotation (movie 3). For the puller with the repulsive boundary condition, the mechanism is similar (movie 4). But due to the absence of direct contacts, only induced asymmetric flows by the heterogeneously distributed polymers can contribute to the rotational enhancement. As a result, it becomes twice weaker. It is worth noting that the monomer density $\rho^m=0.18 a^{-3}$ at the front is higher compared to the no-slip boundary condition case. This is because repulsion is a long-range interaction, causing polymers to be slowed down and accumulate as they approach the squirmer. Due to the same reason, in contrast to the no-slip boundary condition case, polymers in the rear are decelerated before reaching the puller. Thus, the corresponding monomer density $\rho=0.06 a^{-3}$ is lower and the contribution to the enhanced rotational motion by the induced asymmetric flows is further reduced. 

Unlike pushers and pullers, neutral swimmers do not exhibit any rotational enhancement, as shown in Figure Fig.~\ref{Dr_0p8}. Because the source dipole flow field \cite{theers2016modeling} generated by a neutral swimmer has limited influence on collective polymer motion, which is evidently revealed in Fig.~\ref{cylin_0p8}(b). In addition, polymers in the vicinity of the neutral swimmer are advected by the flows and rapidly skim over its surface (movies 5 and 6, Fig.~\ref{Dr_0p8}(a) sketch). Thus, no apparent momentum and angular momentum exchanges occur and the induced asymmetric flows are also weak.

The influence of active stress on the squirmer’s rotational motion is presented in Fig.~\ref{Dr_0p8}(b), where a strong pusher/puller $|\beta|=5$ generates a profound enhancement of $10 \sim 50$ times. For a weak pusher with $\beta=-2$, the polymer wrapping effect is attenuated and the corresponding enhancement is 5 times. For a weak puller with $\beta=2$, the backward suction flow is significantly diminished, leading to no apparent enhancement. In general, the squirmer-monomer no-slip boundary condition yields additional collisions between squirmer and polymers, which generate forces and torques to facilitate rotational motion. The disparity between the two types of boundary conditions is more pronounced when a strong active stress with $|\beta|=5$ is applied. Whereas, the difference is subtle for a squirmer with a weak active stress $|\beta|\le2$, because the generated flows are not sufficiently strong to advect polymers to the vicinity of the squirmer. Consequently, the contribution from the polymer-squirmer collisions is absent.

\begin{figure}
  \centering
  \includegraphics[width=\columnwidth]{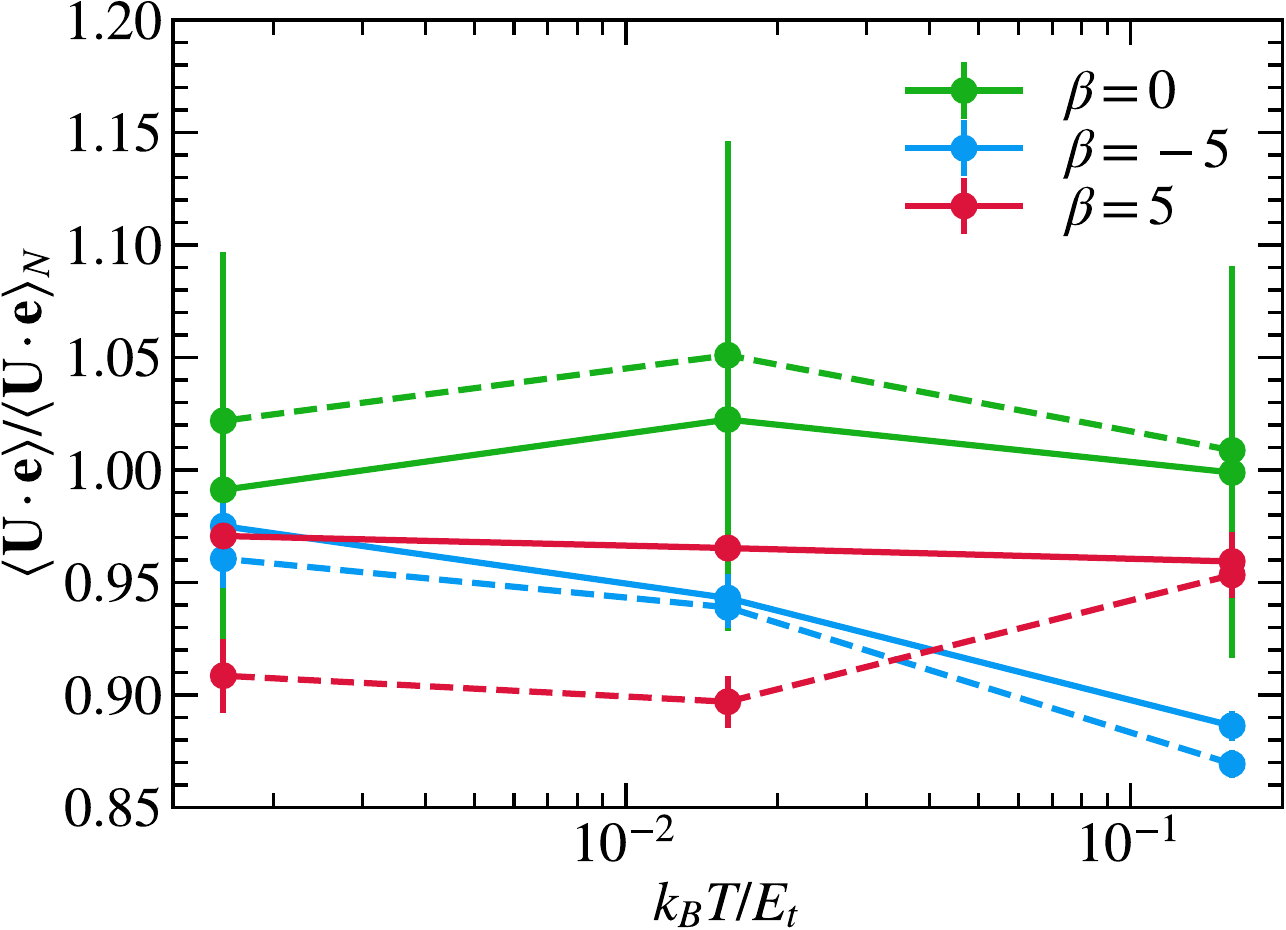}
    \caption{
        Dependences of squirmers' mean swimming velocities on the temperature at ${\rm Re}=0.8$ in the viscoelastic
        fluids. Dashed/solid lines indicate the no-slip/repulsive boundary condition case. Velocities are normalized
        by the Newtonian counterparts $\langle {\bf U} \cdot {\bf e} \rangle_N$. 
    }
  \label{Ue_0p8}
\end{figure}

The mean swimming velocities $\langle {\bf U} \cdot {\bf e} \rangle$ for various swimmers are demonstrated in Fig.~\ref{Ue_0p8}. The combination of the finite Reynolds number and the different self-propulsion schemes results in distinct behavior among the swimmers. For neutral swimmers with the repulsive boundary condition, the mean swimming velocities in the polymer solution are very close to the Newtonian case. This is because the source dipole flow field generated by a neutral swimmer can quickly advect the neighboring polymers to the rear, leaving minimal impact on its translation. Counterintuitively, when the no-slip boundary condition is applied, polymers can get closer to the vicinity of the swimmer, but a $\sim 5\%$ enhancement in the mean swimming velocity is observed. However, considering the large fluctuations in $\langle {\bf U} \cdot {\bf e} \rangle$, it is important to note that this enhancement should not be overemphasized. In the case of pullers, the influence of polymers is more pronounced. The force dipole flow field generated by the puller results in numerous monomer-squirmer collisions at the front, which impede its translation. Additionally, the backwards suction polymers further slow down its motion. When the repulsive boundary condition is applied, the slowing down effect is approximately $4\%$.  In contrast, it is $10\%$ in the no-slip boundary condition case due to the close contact between polymers and pullers at low temperatures. However, this difference diminishes at $k_BT/E_t=0.15$. It is consistent with the change in the rotational enhancements of pullers in Fig.~\ref{Dr_0p8}, where the thermal effect eliminates the discrepancy caused by different boundary conditions at high temperatures. For pushers, the distinction between the two boundary conditions is subtle, i.e., the no-slip boundary condition only leads to an additional $1.5\%$ reduction in $\langle {\bf U} \cdot {\bf e} \rangle$ compared to the repulsive case. Moreover, the thermal effect is more severe for pushers. Consistent decays in the mean swimming velocities from $3\%$ to $13\%$ are observed as temperature increases. 


\subsection{Transport and rotational properties at ${\rm Re}=0.04$}

\begin{figure}
  \centering
  \includegraphics[width=\columnwidth]{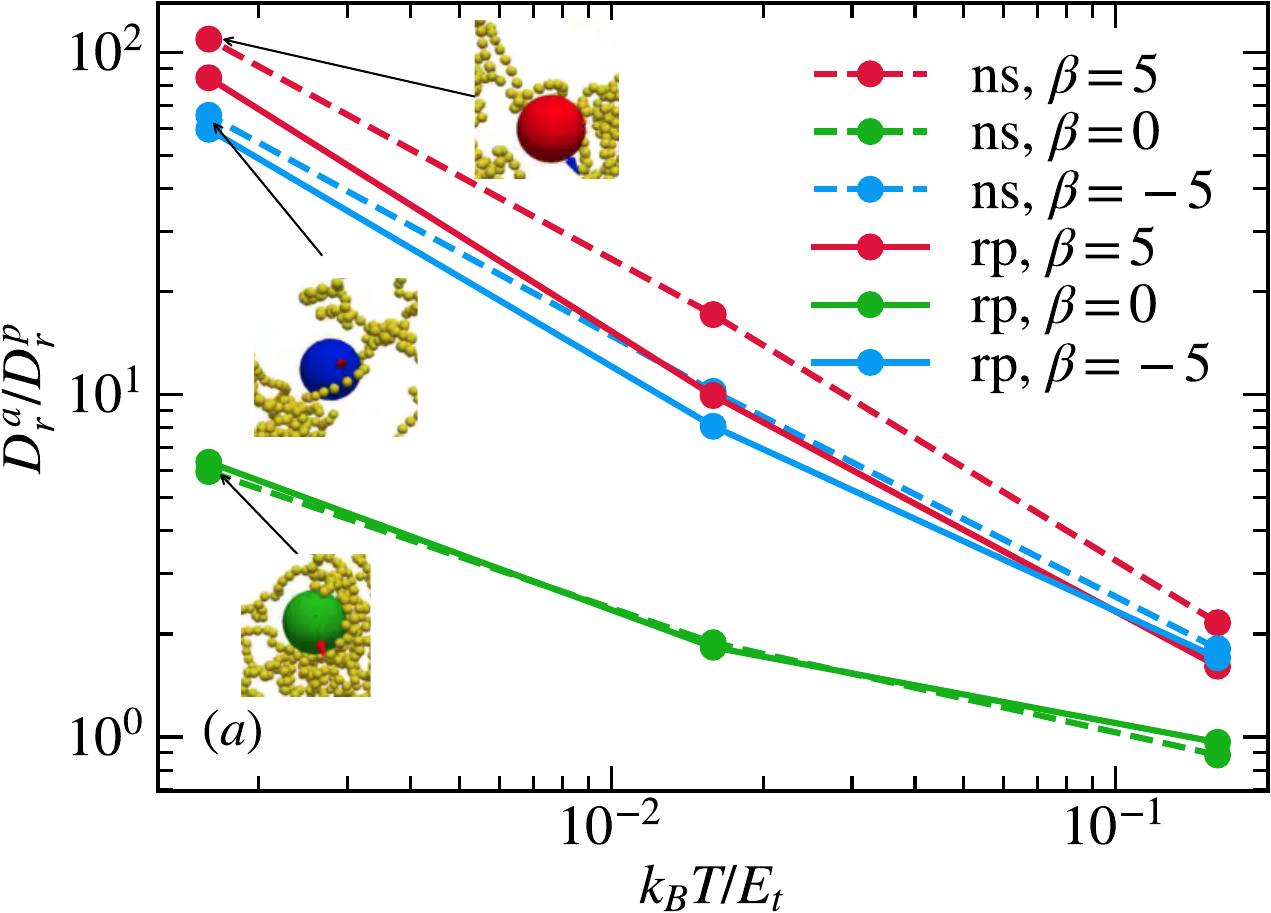}
  \includegraphics[width=\columnwidth]{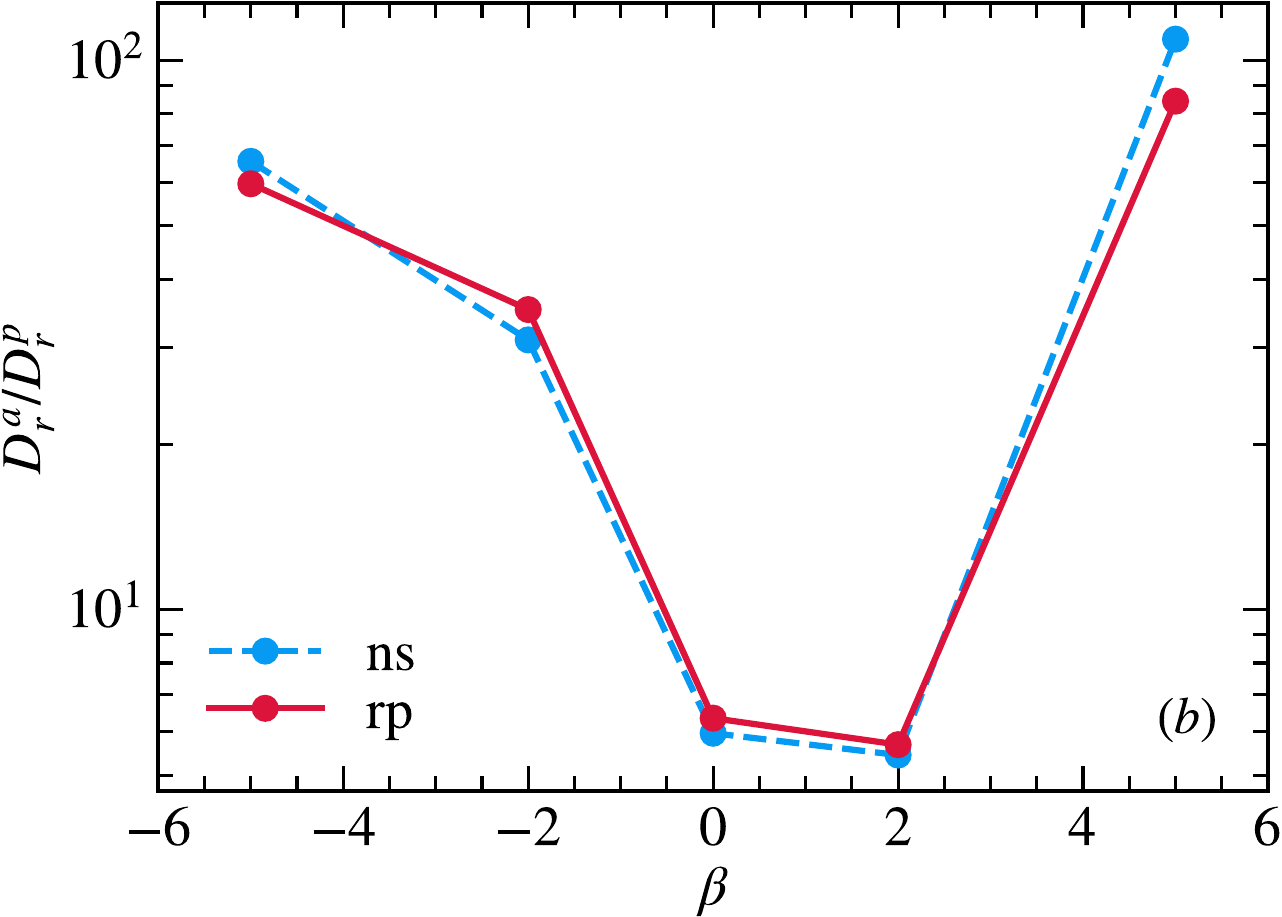}
    \caption{
        (a) Rotational diffusion enhancements for swimmers with the no-slip (ns) and the repulsive (rp)
        boundary conditions at ${\rm Re}=0.04$. Here, the rotational diffusion coefficients for passive colloids
        with the no-slip boundary condition 
        are $D^{p,ns}_r=4.1\times 10^{-9}, 3.5\times 10^{-8}, 3.8\times 10^{-7} \ \tau^{-1}$
        at temperatures $k_BT/E_t=0.0015, 0.015, 0.15$, respectively.
        Similarly, $D^{p,rp}_r=4.0\times 10^{-9}, 3.9\times 10^{-8}, 4.0\times 10^{-7} \ \tau^{-1}$ for
        the case with the repulsive boundary condition. Snapshots are the representative configurations 
        for squirmers with the no-slip boundary condition at $k_BT/E_t=0.15$. (b) Dependences of rotational 
        enhancements on the active stress at $k_BT/E_t=0.0015$.
    }
  \label{Dr_0p04}
\end{figure}

\begin{figure*}
  \centering
  \includegraphics[width=0.8\paperwidth]{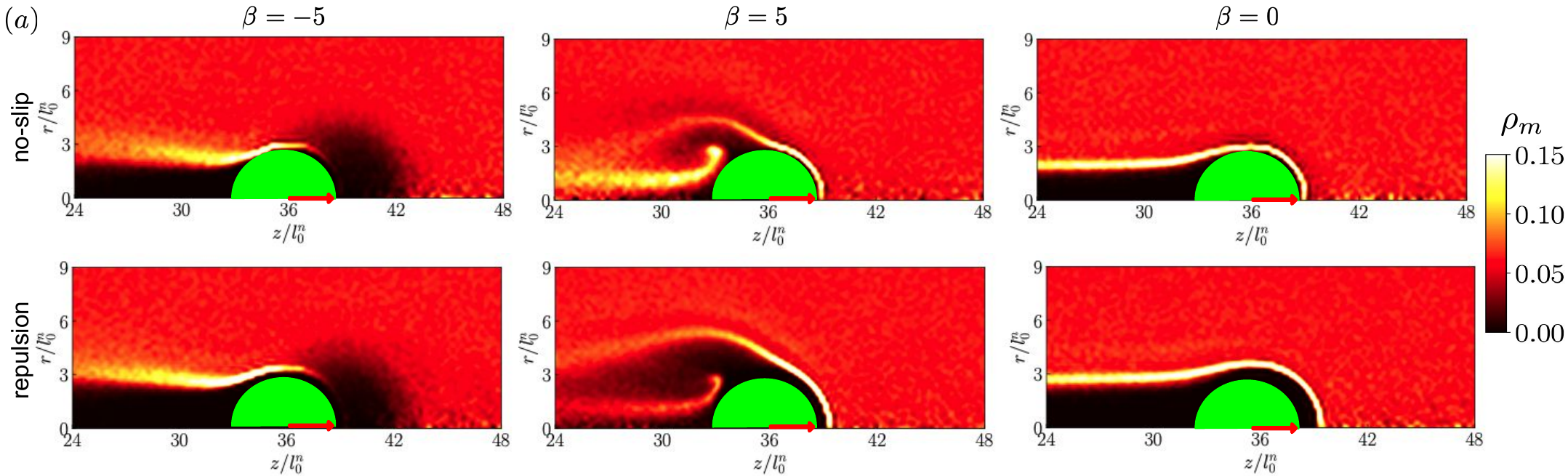}
  \includegraphics[width=0.8\paperwidth]{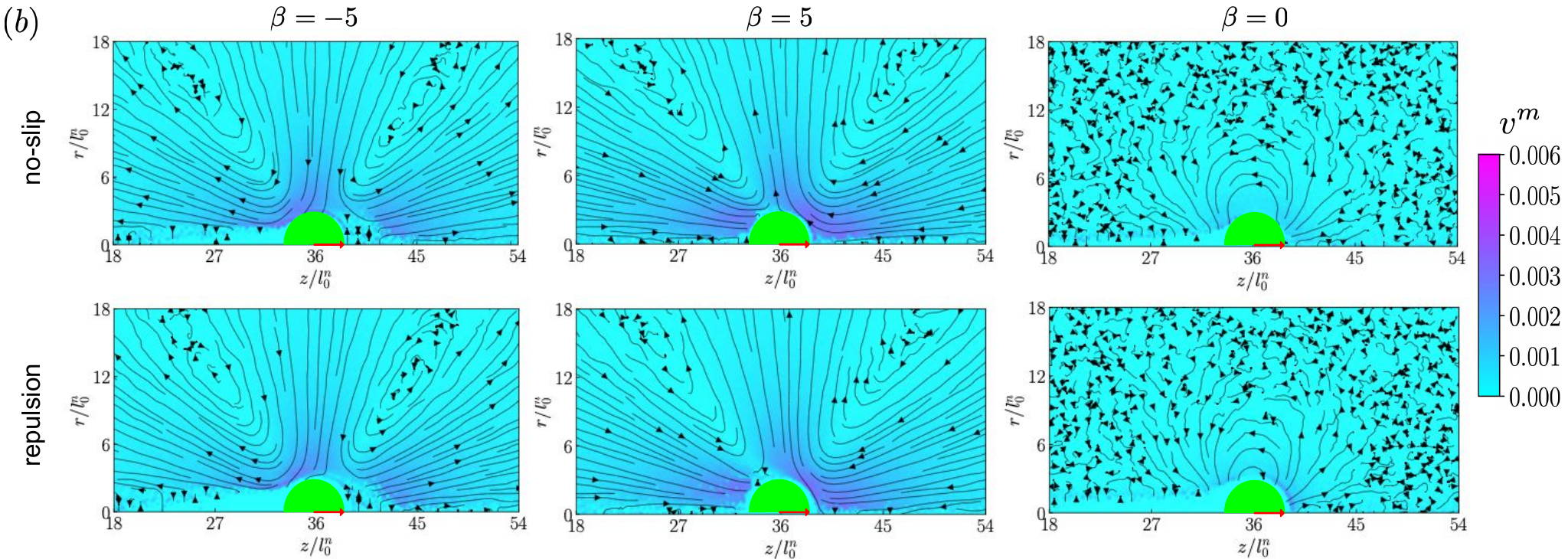}
    \caption{
        Cylindrical monomer (a) density and (b) velocity distributions of pushers ($\beta=-5$), 
        pullers ($\beta=5$), and neutral squirmers ($\beta=0$) at ${\rm Re}=0.04$. Here, the no-slip
        and repulsive boundary conditions are applied, respectively.}
  \label{cylin_0p04}
\end{figure*}

Similar to the previous case, the rotational enhancements for swimmers at ${\rm Re}=0.04$ also exhibit temperature dependencies. As illustrated in Fig.~\ref{Dr_0p04}, the interplay of activity and the system heterogeneity determines the enhancements at low temperatures, but the thermal effect quickly becomes dominant at $k_BT/E_t = 0.15$. Strikingly, the enhancements at ${\rm Re}=0.04$ are more pronounced than the ones at the finite Reynolds number. In particular, reminiscent of the experiments where a Janus particle is immersed in viscoelastic solutions \cite{gomez2016dynamics}, the enhancement for a puller with the no-slip boundary condition exceeds two orders of magnitude. However, such a strong enhancement in our case is primarily attributed to the significant reduction of the rotational diffusion of a passive colloid in the solution with the high viscosity $\eta=0.5 ma/\tau$, i.e., $D^p_r=4\times 10^{-9} \tau^{-1}$ at $k_BT/E_t = 0.0015$. The interplay of activity and the system heterogeneity in our fluid-like polymer solutions can hardly lead to the circular motion as observed in experiments, which is attributed to the memory effect of viscoelastic fluids \cite{narinder2018memory}. This effect originates from the misalignment between the instantaneous particle orientation and the internal forces generated by the deformed polymers during self-propulsion. However, our polymer solutions are more fluid-like, the observed enhancements in the rotational diffusion here are mainly due to the squirmer-polymer mechanical couplings. Contrary to the ${\rm Re}=0.8$ case, the no-slip boundary condition does not have a profound extra contribution to the enhancement (Fig.~\ref{Dr_0p04}(a)). Because viscous drag is dominant at the low Reynolds number, the polymer's motility is strongly suppressed. Consequently, it becomes more challenging for the polymers to reach the squirmers, leading to a large reduction in momentum and angular momentum exchanges. This phenomenon is evident in Fig.~\ref{cylin_0p04}(a), where the monomer densities in the vicinity of squirmers are generally smaller compared to those in Fig.~\ref{cylin_0p8}(a).

As shown in Fig.~\ref{cylin_0p04}, the decrease in the polymer motility results in an apparent reduction in the monomer density near a pusher. Especially, it approaches zero in front of the squirmer. Thus, the polymer wrapping effect is strongly suppressed. However, as demonstrated in movie 7 where a no-slip boundary condition is applied between polymers and the pusher, a considerable amount of polymer wrappings still exist. They generate additional momentum and angular momentum exchanges to facilitate the squirmer's rotation. But their influences are less substantial as at ${\rm Re}=0.8$. Compared with the repulsive case, the relative enhancement in Fig.~\ref{Dr_0p04}(b) is only 1.2, even though the net value is 66. 

The influence of decreased polymer motility is less severe for pullers. This is because the mechanism involving the heterogeneously suctioned polymers in the rear of the puller is retained (movies 9, 10, Fig.~\ref{Dr_0p04}(a) sketch, and Fig.~\ref{cylin_0p04}(a)). Thus, in Fig.~\ref{Dr_0p04}(a), a significant net enhancement over 110 times is observed in the no-slip boundary condition case at $k_BT/E_t = 0.0015$, which is $1.3$ times stronger compared to the repulsive case. This is because the reduced motility inhibits polymers from reaching the squirmer. Consequently, momentum and angular momentum exchanges under the no-slip boundary condition are suppressed. This phenomenon is revealed in Fig.~\ref{cylin_0p04}(a), where monomer concentration in the rear of the squirmer becomes lower and narrower in the no-slip boundary condition case. It is worth addressing that the broadness of the monomer density distribution is important. Because the further the polymers are away from the centerline, the stronger the torques generated by the asymmetric flows are. In the repulsive boundary condition case, the distribution becomes brighter but narrower compared to the case at ${\rm Re}=0.8$. Here, polymers are closer to the centerline and their influences on the rotational enhancement of the puller are less profound (movies 4 and 10). 

Similar to the case at ${\rm Re}=0.8$, neutral swimmers also do not exhibit profound rotational enhancements due to the rapid polymer advection process (movies 11 and 12, Fig.~\ref{Dr_0p04}(a) sketch). A 6 times enhancement is found at $k_BT/E_t = 0.0015$ in Fig.~\ref{Dr_0p04}(a). 

As shown in Fig.~\ref{Dr_0p04}, the influence of active stress on the rotational enhancement at ${\rm Re}=0.04$ is similar to that in the finite Reynolds number regime. To be more specific, strong enhancements around $60\sim100$ times are found for $|\beta|=5$. A pusher with $\beta=-2$ exhibits 30 times enhancement. But for a weak puller ($\beta=2$) and a neutral swimmer, the enhancements are only 5 times. The significant change for a puller when decreasing the active stress is due to the reduction of absorbed polymers in the rear of the squirmer. Thus, the asymmetric flows induced by the heterogeneous distributed polymers are largely suppressed. Since fluid viscous drag greatly reduces the polymer's motility at ${\rm Re}=0.04$, the additional enhancements caused by the monomer-squirmer no-slip boundary condition are small.  

It is noteworthy that the enhanced rotational motion of squirmers is entirely attributed to the intrinsic properties of polymers. For comparison, the rotational diffusion of squirmers in monomer fluids is also investigated, with the monomer concentration kept consistent with previous study. The rotational diffusion coefficients for the pusher, puller, and neutral swimmer are $D^m_r=2.4, 6.7, 8.0 \times 10^{-9} \tau^{-1}$, respectively, which are close to the theoretical estimation of $3.0 \times 10^{-9} \tau^{-1}$ for a spherical squirmer in Newtonian fluid. 

\begin{figure}
  \centering
  \includegraphics[width=\columnwidth]{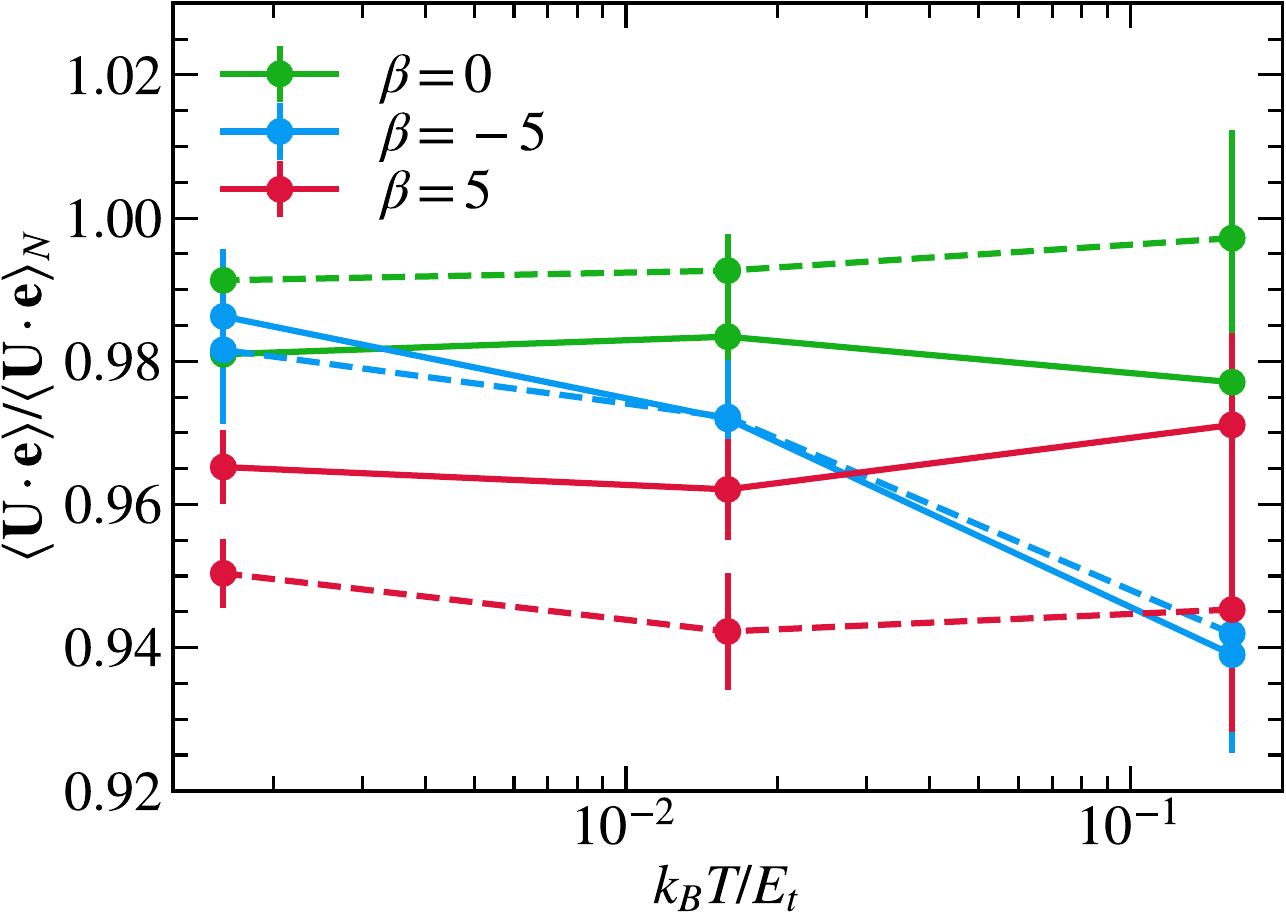}
    \caption{
        Dependences of squirmers' mean swimming velocities on the temperature at ${\rm Re}=0.04$ in the viscoelastic fluids. Dashed/solid lines indicate the no-slip/repulsive boundary condition case. Velocities are normalized
        by the Newtonian counterparts $\langle {\bf U} \cdot {\bf e} \rangle_N$. 
    }
  \label{Ue_0p04}
\end{figure}

The transport properties of swimmers at ${\rm Re}=0.04$ are presented in Fig.~\ref{Ue_0p04}. Consistent with previous studies \cite{zhu2012self,binagia2020swimming}, the fluid viscoelasticity leads to reductions in the mean swimming velocities for all the squirmers compared to the Newtonian counterparts.  For neutral swimmers, due to the rapid polymer advection effect, the reduction is only $1\%$ in the case with the no-slip boundary condition and $2\%$ when the repulsion is applied. Front monomer-squirmer collisions and backwards suction polymers lead to stronger reductions in pullers, which are $3.5\%$ and $5\%$ in the cases with the repulsive and the no-slip boundary conditions, respectively. The latter is more profound because polymers can get closer to the puller. Consistent decays from $2\%$ to $6\%$ in the mean swimming velocities of pushers are also found at ${\rm Re}=0.04$. Similarly, the differences between the two boundary conditions diminish due to the reduction of polymer motility, resulting in fewer monomer-squirmer collisions. 


\subsection{Further discussions at ${\rm Re}=0.04$}

The influences of monomer packing fraction, polymer length, and squirmer-polymer aspect ratio on the rotational and translational behavior of squirmers at ${\rm Re}=0.04$ are addressed below. Given that the extra contributions stemming from the no-slip boundary condition are small, only the repulsive case is considered. 

\begin{figure}
  \centering
  \includegraphics[width=\columnwidth]{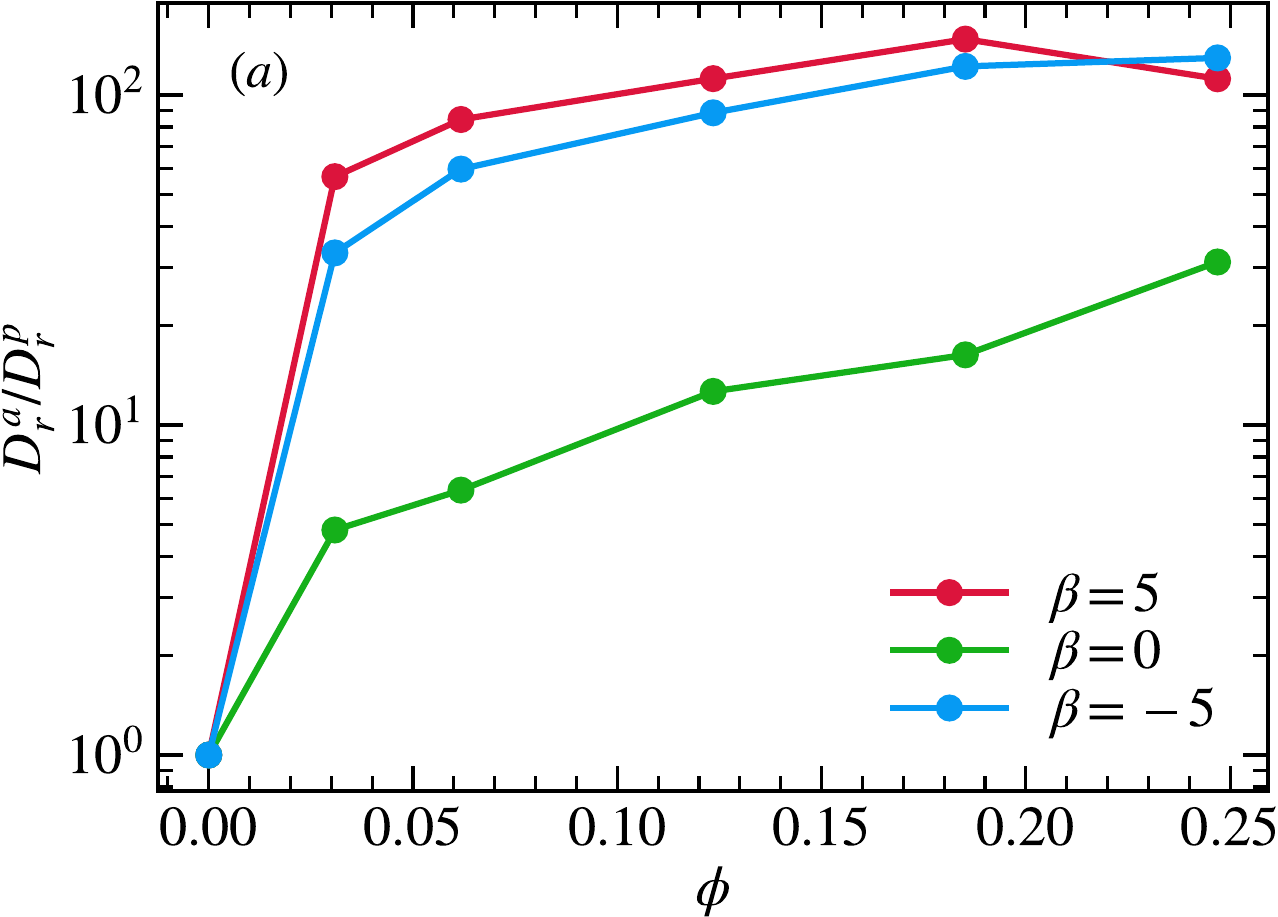}
  \includegraphics[width=\columnwidth]{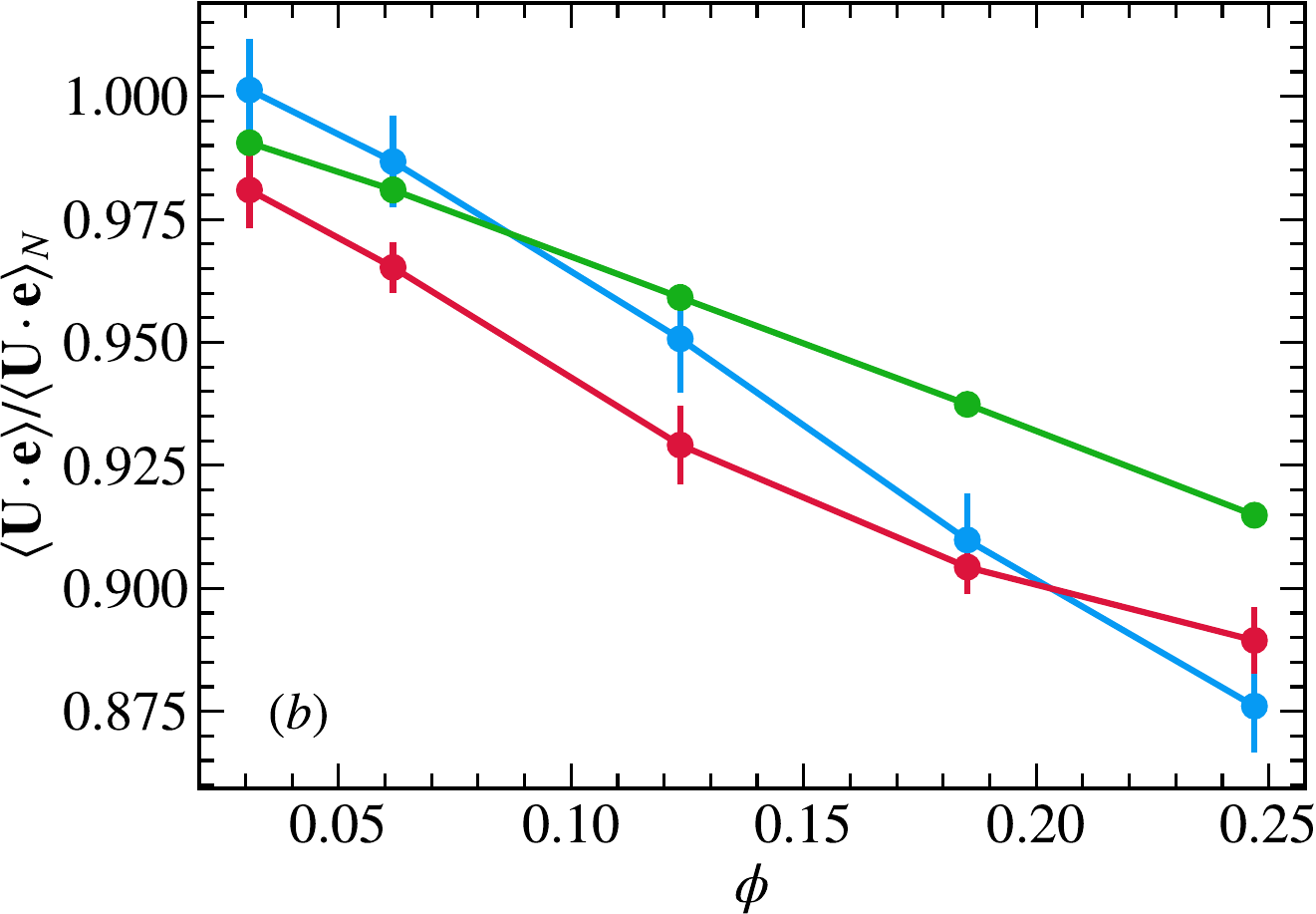}
    \caption{
    (a) Dependences of the rotational enhancements $D^a_r/D^p_r$ on the monomer packing fraction $\phi$
    at $k_BT/E_t=0.0015$. The rotational motion of a passive colloid is intact when raising $\phi$.
    The corresponding rotational diffusion coefficient is $D^p_r=3.0\times 10^{-9} \ \tau^{-1}$. 
    (b) Dependences of normalized mean swimming velocities on the monomer packing fraction.
    }
  \label{Dr_Ue_phi}
\end{figure}

The dependences of rotational enhancements for squirmers with different active stresses on the monomer packing fraction are exhibited in Fig.~\ref{Dr_Ue_phi}(a). In the absence of polymers, a homogeneous system is obtained, resulting in squirmers with various active stresses possessing identical rotational diffusion similar to a passive colloid. The corresponding rotational enhancements are unity. However, they increase rapidly to $33$ for a pusher and $56$ for a puller when the packing fraction raises to $\phi=0.03 a^{-3}$. A $5$ times enhancement is observed for a neutral swimmer. These enhancements are attributed to the increase in the system's heterogeneity caused by the embedded polymers \cite{qi2020enhanced}. They induce asymmetric
flows in the vicinity of swimmers and thus facilitate their rotational diffusion. The enhancements become more significant at $\phi=0.19 a^{-3}$, reaching 122 times for the pusher, 148 times for the puller, and 16 times for the neutral swimmer. Afterwards, $D^a_r/D^p_r$ reaches a saturated value for the pusher, but it starts to decrease for the puller, indicating the vanishing of the asymmetric flows induced by polymers and the transition of the system to a more homogeneous phase \cite{qi2020enhanced}. 

The influences of monomer packing fraction $\phi$ on the mean swimming velocities are addressed in Fig.~\ref{Dr_Ue_phi}(b). An increase in polymer concentration enlarges the viscous resistance, which further impairs the transport of squirmers. However, the impact has a dependence on active stress. For a neutral swimmer, due to the rapid advection of polymers on its surface, the reduction in the mean swimming velocity is $8\%$ at $\phi=0.25 a^{-3}$. But it becomes $11\%$ for a puller. This is because numerous polymer-squirmer collisions at the front and the backwards suction polymers cooperatively slow down the translation of the puller. For a pusher, the mean swimming velocity is closer to its Newtonian counterpart at low monomer packing fractions. But the polymer effect becomes more evident as $\phi$ increases. The velocity reduction reaches $12\%$ at $\phi=0.25 a^{-3}$. This is, on the one hand, due to the enhanced rotation stemming from the wrapped polymers at the front. On the other hand, a large amount of monomer-squirmer collisions on the side of the pusher also interfere with its translation.

\begin{figure}
  \centering
  \includegraphics[width=\columnwidth]{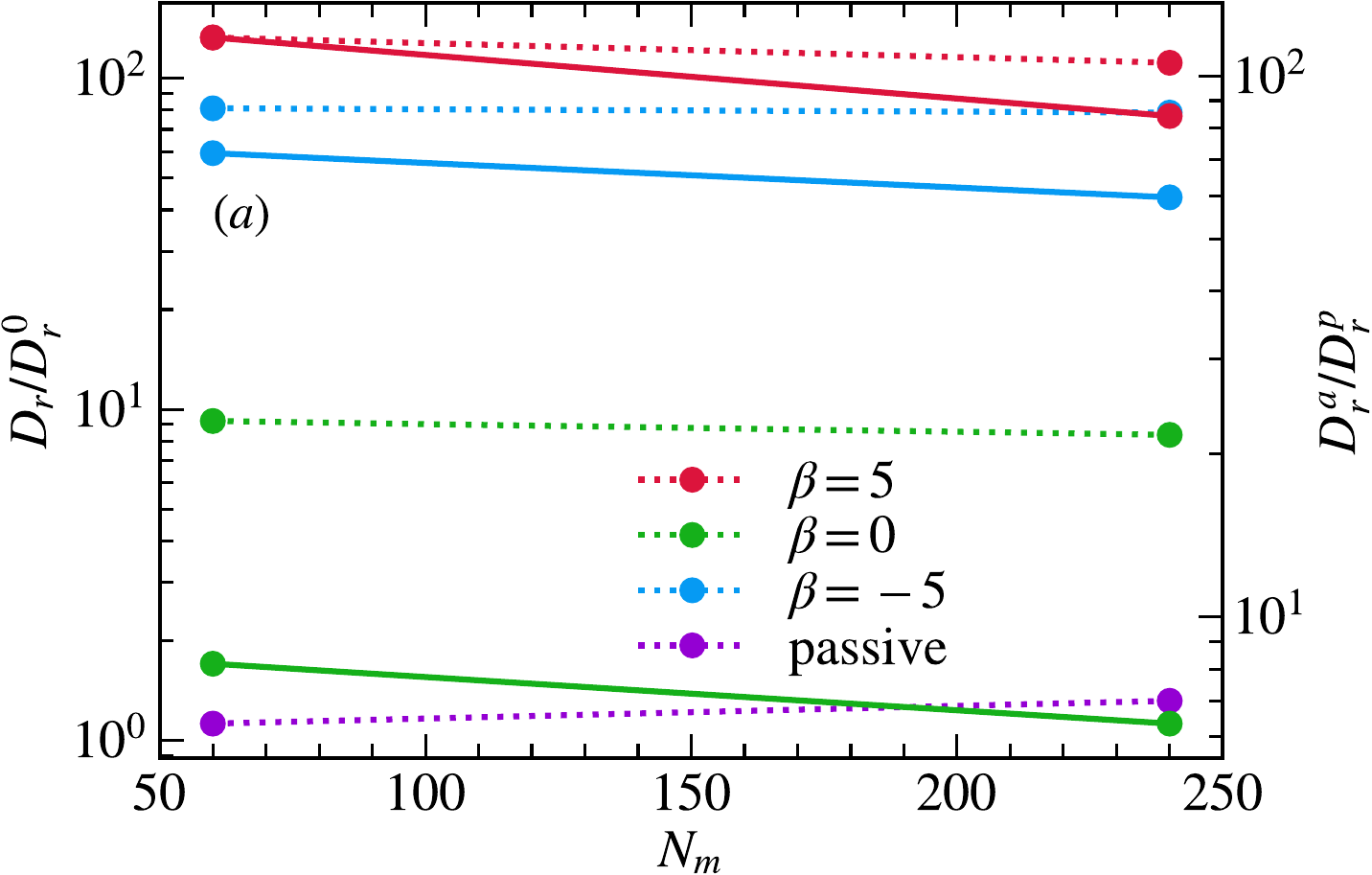}
  \includegraphics[width=\columnwidth]{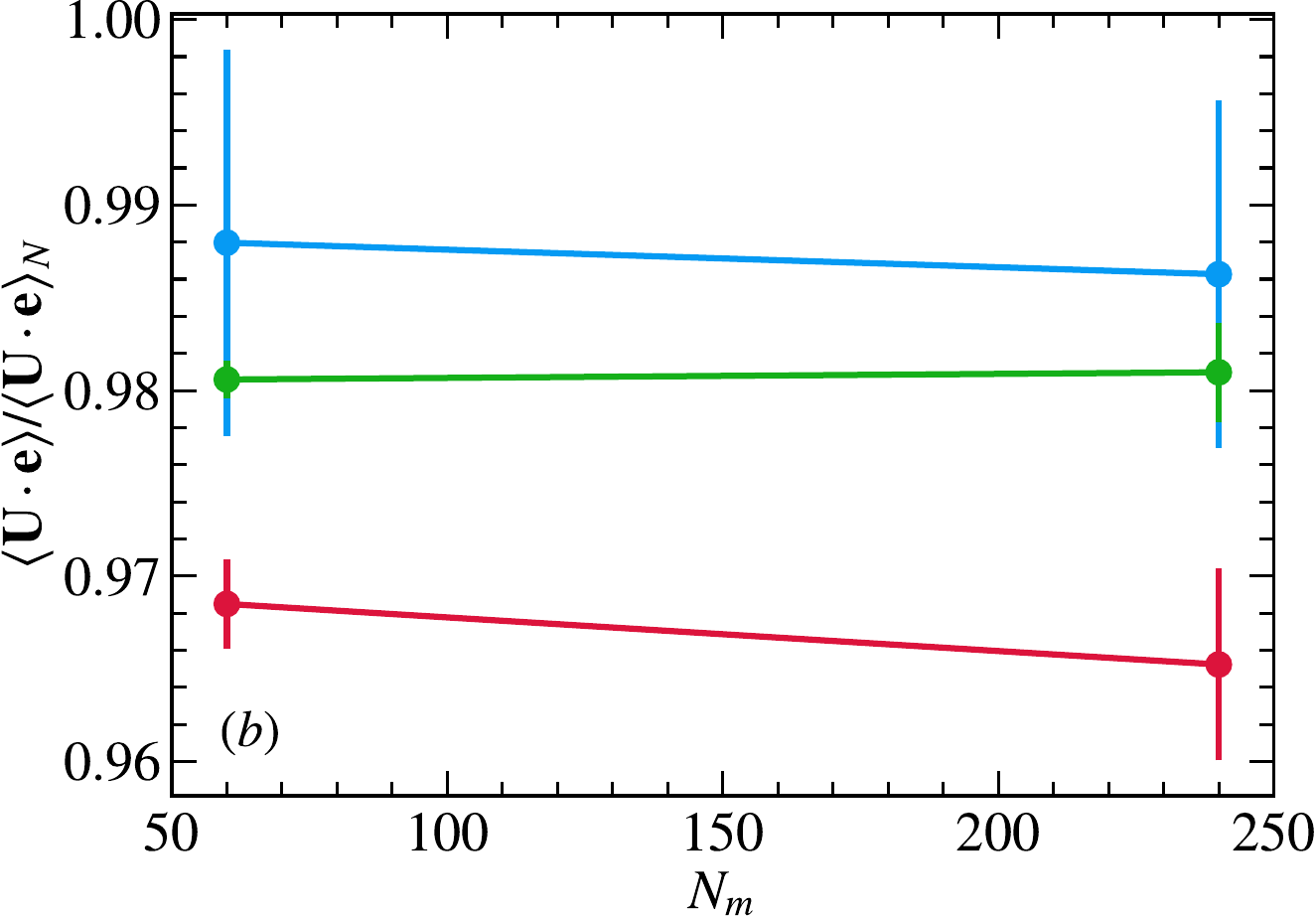}
    \caption{
    (a) (Dotted lines) Dependences of rotational diffusion coefficients $D_r/D^0_r$ 
    on the polymer length $N_m$ for a passive colloid and squirmers at ${\rm Re}=0.04$. (Solid lines)  
    Dependences of rotational enhancements $D^a_r/D^p_r$ on the polymer length. Here, the repulsive boundary condition 
    is applied. Monomer packing fraction $\phi=0.06 a^{-3}$ is fixed for both polymer lengths. (b) Dependences of 
    normalized mean swimming velocities on the polymer length.
    }
  \label{Dr_Ue_poly_length}
\end{figure}

By fixing monomer packing fraction $\phi=0.06 a^{-3}$, the influences of polymer length on the squirmer's rotational enhancements are investigated in Fig.~\ref{Dr_Ue_poly_length}(a). The enhancements increase by nearly $35\%$ when shorter polymers with $N_m=60$ are employed. This small increase originates mainly from the decreased rotation diffusion of the passive colloid. However, rotational enhancement is a sensitive quantity. A small fluctuation in measuring $D_r$ will be amplified in $D^a_r/D^p_r$. Thus, the small increase should not be overemphasized. In \cite{qi2020enhanced}, the rotational enhancement decreases twice when shorter polymers are used. This is mainly due to the large decrease of $D_r$ of a passive colloid. In that work, a short-range attraction is applied between a squirmer and polymers, resulting in strong polymer absorption on the squirmer's surface, which greatly suppressed its rotational diffusion. In contrast, the repulsive boundary condition is used in our model. Thus, the rotational diffusion of a passive colloid is almost intact when varying the polymer length.

As shown in Fig.~\ref{Dr_Ue_poly_length}(b), similar to the rotational motion, an increase in polymer length has no evident influence on the mean swimming velocities of squirmers. Compared with the Newtonian counterparts, the reductions in $\langle {\bf U} \cdot {\bf e}\rangle$ are $1.4\%$, $2\%$, and $3.5\%$ for a pusher, a neutral swimmer, and a puller, respectively.

\begin{figure}
  \centering
  \includegraphics[width=\columnwidth]{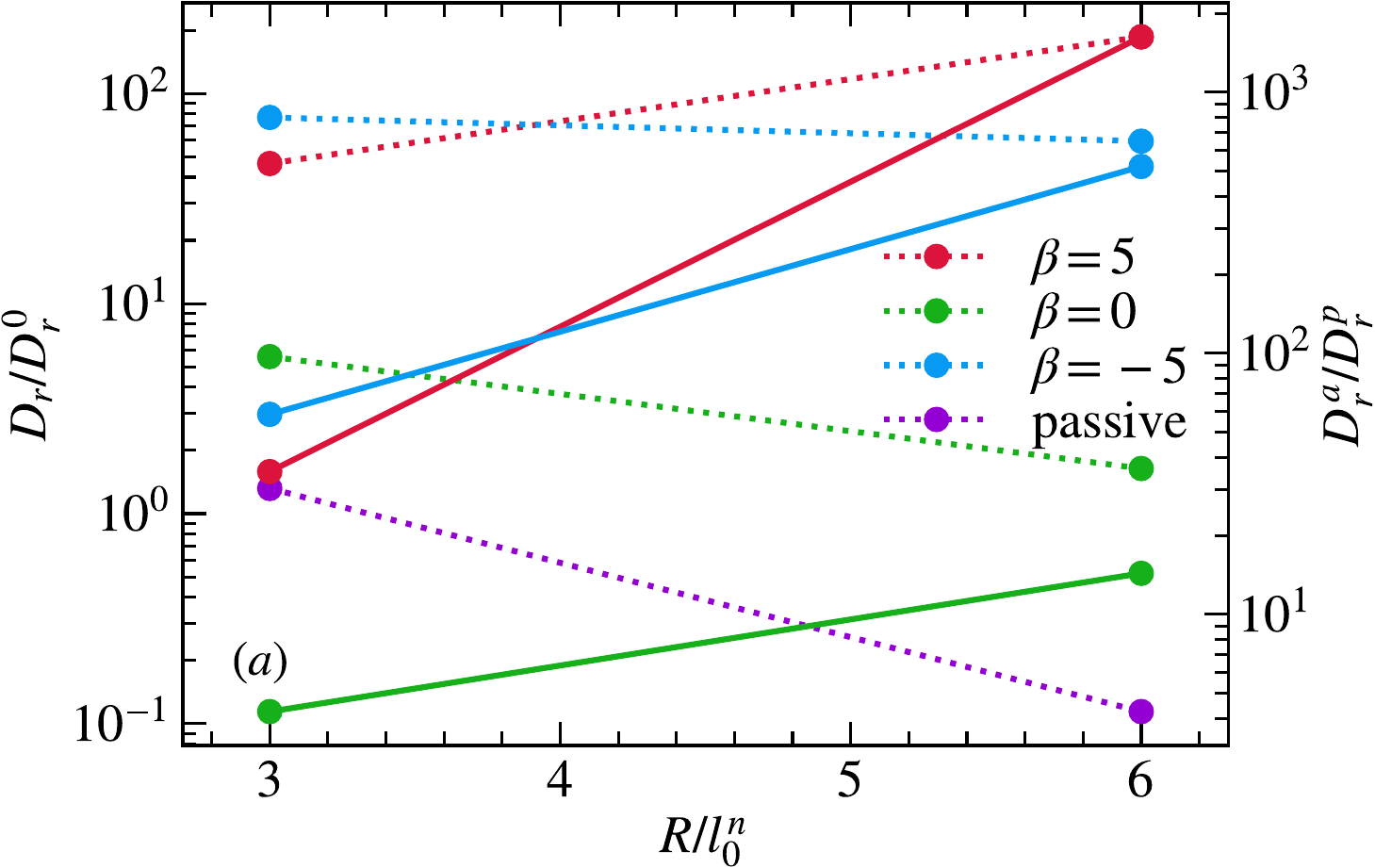}
  \includegraphics[width=0.9\columnwidth]{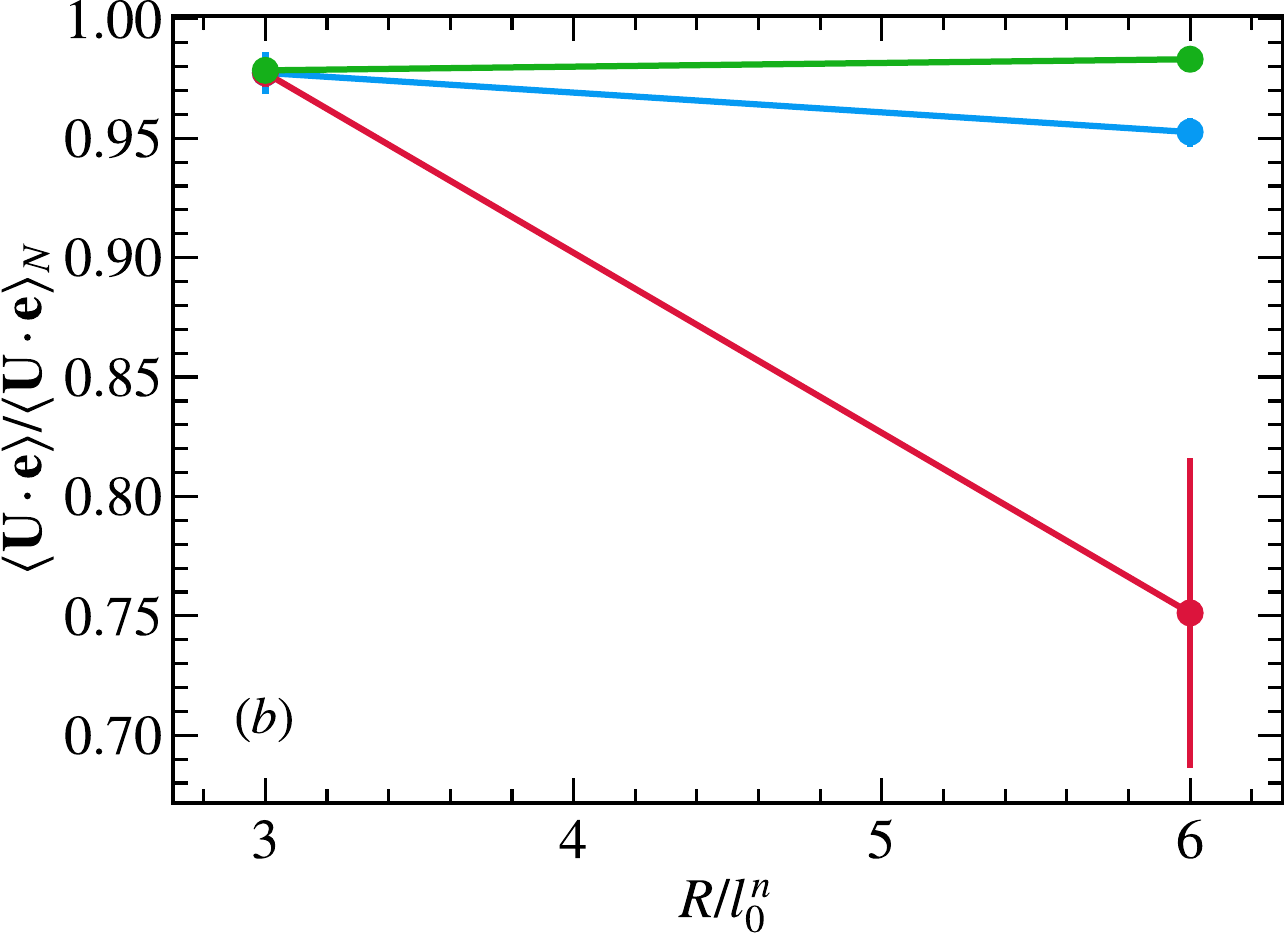}
    \caption{
    (a) (Dotted lines) Dependences of rotational diffusion coefficients $D_r/D^0_r$ of passive colloids and 
    squirmers on the particle radius $R/l^n_0$ at ${\rm Re}=0.04$. 
    (Solid lines) Dependences of rotational enhancements $D^a_r/D^p_r$ on the particle radius. 
    The monomer packing fraction $\phi=0.06 a^{-3}$ is fixed. 
    Radius is normalized by the polymer natural bond length $l^n_0$.
    $D^0_r=3.0\times 10^{-9} \ \tau^{-1}$ is the theoretical estimation of the rotational diffusion 
    coefficient of a colloid with  $R=3a$ at temperature $k_BT/E_t=0.0015$.
    (b) Dependences of normalized mean swimming velocities on the particle size.
    }
  \label{Dr_Ue_aspect_ratio}
\end{figure}

\begin{figure*}
  \centering
  \includegraphics[width=0.6\paperwidth]{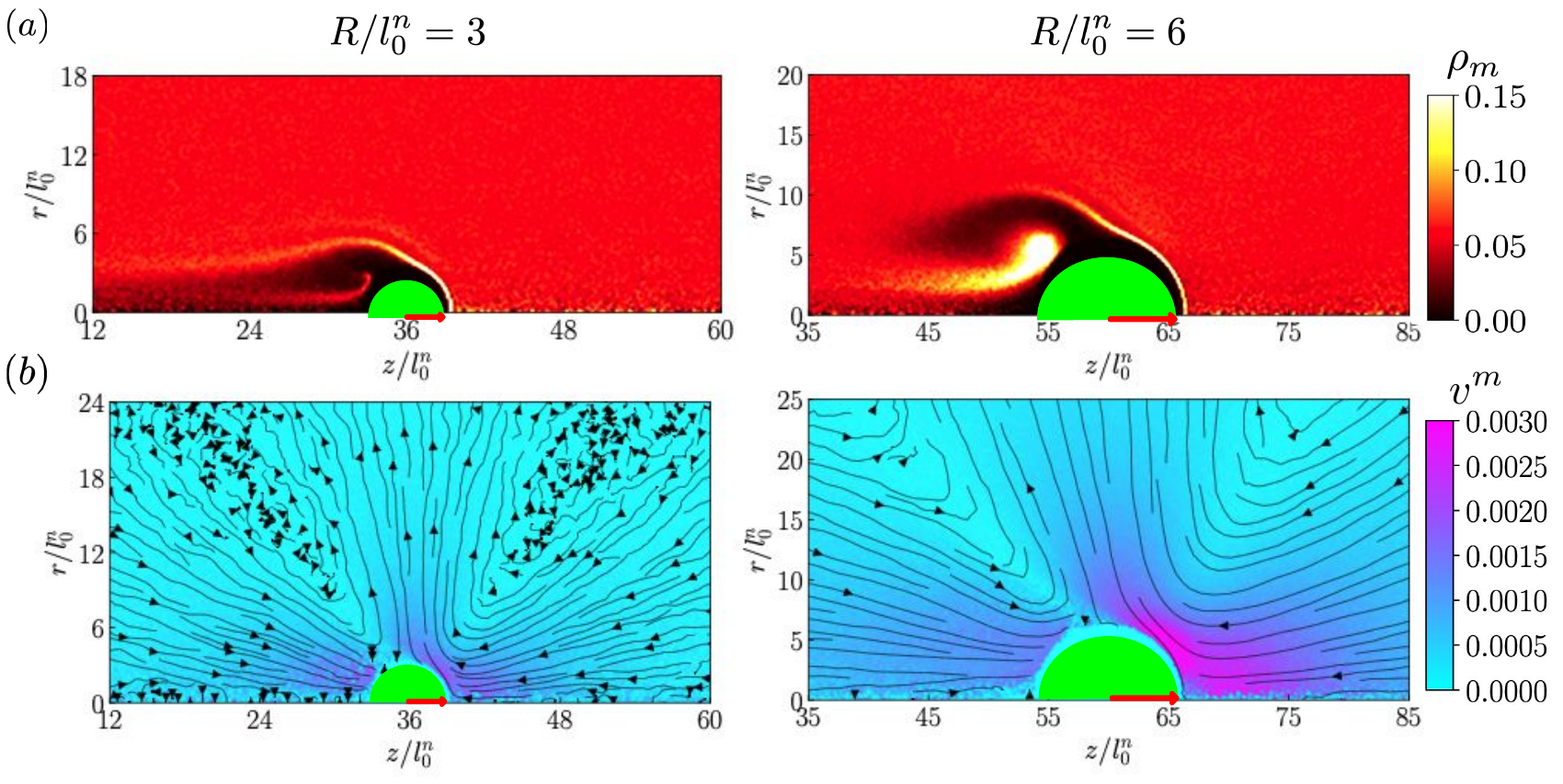}
    \caption{
        Cylindrical monomer (a) density and (b) velocity distributions of pullers ($\beta=5$) 
        with aspect ratio $R/l^n_0=3, 6$ at ${\rm Re}=0.04$. Here, the repulsive boundary condition
        is applied.}
  \label{cylin_0p04_R36}
\end{figure*}

The dependences of rotational enhancements on squirmer-polymer aspect ratio are investigated in Fig.~\ref{Dr_Ue_aspect_ratio}(a). Swimmers with $R=3,6a$ are considered. To prevent excessively strong flows that could break polymer bonds, a smaller $B_1=0.003$ is used. Consequently, the corresponding Reynolds numbers become ${\rm Re}=0.024/0.048$ for $R=3/6a$, ensuring that the mechanisms associated with transport and rotational properties remain intact. As shown in Fig.~\ref{Dr_Ue_aspect_ratio}(a), the rotational diffusion coefficient of a passive colloid obeys the Stokes–Einstein relation and decays about a decade as particle size increases. However, activity can compensate for the intrinsic decay of $D_r$. The rotational diffusion coefficients of a pusher and a neutral swimmer only decrease 2 and 3 times, resulting in rotational motions that are 5 and 3 times stronger. Remarkably, the puller's $D_r$ increases 4 times, leading to a significant enhancement with $D^a_r/D^p_r=1400$. This is attributed to the fact that a larger puller can generate stronger and more stable flows in its rear, which efficiently advect polymers to its vicinity. As a result, a noticeable increase in the monomer density can be observed in Fig.~\ref{cylin_0p04_R36}(a). These polymers induce intense asymmetric flows, thereby tremendously enhancing the rotational motion of the puller (movie 13). 

Moreover, the influences of the squirmer-polymer aspect ratio on the mean swimming velocities are demonstrated in Fig.~\ref{Dr_Ue_aspect_ratio}(b). Compared to the Newtonian counterparts, an increase in squirmer size has no apparent impact on the mean swimming velocities of a pusher and a neutral swimmer. In contrast, a large decay of $35\%$ in $\langle {\bf U} \cdot {\bf e} \rangle/\langle {\bf U} \cdot {\bf e} \rangle_N$ is observed for a puller with $R/l^n_0=6$. This decay is primarily caused by the strong hindrance imposed by the backwards suction polymers. These polymers induce intense asymmetric flows in the rear of the puller, leading to a significant enhancement in its rotational motion and, consequently, impeding its transport.

\section{Summary and conclusions}

In our study, we conducted comprehensive investigations into the transport and rotational properties of squirmers with different active stresses in polymer solutions using mesoscale hydrodynamic simulations, i.e., the Lattice-Boltzmann method. To thoroughly elucidate the relevances of fluid viscoelasticity, we have investigated the swimming behavior of swimmers at finite and low Reynolds numbers. 

At ${\rm Re}=0.8$, fluid inertia is dominant. The flows stemming from swimming can readily advect polymers to the vicinities of squirmers. This process promotes collisions between the swimmers and polymers, when the no-slip boundary condition is applied. As a consequence, a pusher has a 50 times enhancement in its rotational motion at $k_BT/E_t=0.0015$, which is 4.2 times stronger compared to the repulsive boundary condition case. This notable increase is primarily attributed to the polymer wrapping at the front of the pusher, which generates asymmetric torques facilitating its rotation. On the other hand, the additional enhancement caused by the no-slip boundary condition is less pronounced for a puller. It can create suction flows in its rear, effectively absorbing polymers and inducing asymmetric flows that accelerate the rotation. The corresponding rotational enhancements for a puller are 26 and 13 in the no-slip and repulsive boundary condition cases, respectively. Since source dipole flow fields rapidly advect polymers to the rear of a neutral swimmer, its rotational motion only increases twice in both cases. As the system temperature rises, the thermal effect becomes dominant, leading to the disappearance of the aforementioned mechanisms. At $k_BT/E_t=0.15$, the rotational diffusion coefficients of squirmers with various active stresses obey the Stokes–Einstein relation. In terms of transport properties, decreases in swimming velocities have been observed in both pullers and pushers. The presence of polymers not only increases the viscous drag of the fluid but also enhances the swimmers’ rotation. The combined effects lead to a slowdown in the translational motion. When the no-slip boundary condition is applied, collisions between squirmers and polymers further accelerate the rotational motion of swimmers. Therefore, compared to the repulsive case, an additional $6\%/2\%$ decrease in the mean swimming velocity has been observed in a puller/pusher at $k_BT/E_t=0.0015$. In contrast, polymers have no severe impact on the translation of neutral swimmers, and their swimming speeds remain close to their Newtonian counterparts. It is worth noting that although the no-slip boundary condition leads to $\sim 5\%$ increase in the swimming speeds of neutral swimmers, this enhancement should not be overemphasized considering the large fluctuations in $\langle {\bf U} \cdot {\bf e} \rangle/\langle {\bf U} \cdot {\bf e} \rangle_N$.

At ${\rm Re}=0.04$, the viscous force of the fluid plays the leading role in determining the behavior of squirmers in polymer solutions. Consequently, the polymer's motility is largely suppressed and the corresponding squirmer-monomer collisions are limited. As a result, the discrepancy between the two boundary conditions is small. In particular, the reduction of polymer motility has less impact on the puller, as the backwards suction polymers are retained to a large extent. They can induce sufficient asymmetric flows to accelerate the rotational motion of the puller, resulting in a profound enhancement of 110 times. Whereas, it turns into 60 for a pusher and 6 for a neutral swimmer at $k_BT/E_t=0.0015$. The translational motion of all types of swimmers is diminished by $1\% \sim 6\%$. Due to the substantial enhancements of the rotational motion, pullers exhibit stronger reductions in swimming velocities. Afterwards, the influences of polymer concentration, polymer length, and squirmer-polymer aspect ratio are addressed, respectively. In short, increasing polymer concentration results in enhancements in squirmers' rotations as systems become more heterogeneous. But the translational motion is largely suppressed due to the impedance caused by polymers. Whereas, the transport and rotational behavior of swimmers have weak dependences on the polymer length. In contrast, increasing the squirmer size can substantially change its behavior. According to the Stokes–Einstein relation, the rotational diffusion coefficient of a passive colloid reduces 8 times as the squirmer's radius doubles. As a consequence, the rotational enhancements $D^a_r/D^p_r$ reach 520 and 15 for a pusher and a neutral swimmer with $R/l^n_0=6$, respectively. Strikingly, $D^a_r/D^p_r \approx 1600$ has been observed for a large puller. This tremendous enhancement stems from the stark backwards suction flows generated by the swimmer, which advect abundant polymers to the rear of the puller and significantly increase its rotation by inducing asymmetric flows. 

In experiments where a Janus particle was immersed in fluids embedded with PAAm polymers, a significant enhancement of rotational diffusion over 400 times was reported \cite{gomez2016dynamics}. In that work, this phenomenon originates from the memory effect of the viscoelastic fluids, namely, the internal forces generated by the deformed polymers during self-propulsion misalign with the instantaneous particle orientation. Further increasing activity even leads to circular motion \cite{narinder2018memory}. For the purpose of comparison, we estimate the dimensionless numbers as followings. By using the water-PnP mixture fluid density $\rho \approx 0.95 \times 10^3 {\rm kg/m^3}$, particle diameter $2R=7.75 {\rm \mu m}$, typical swimming speed $U_0=0.25 {\rm \mu m /s}$, and fluid viscosity $\eta=0.15 {\rm Pa \, s}$ \cite{gomez2016dynamics,narinder2018memory}, the system Reynolds number can be calculated as ${\rm Re}=2R\rho U_0/\eta=1.3 \times 10^{-8}$. The fluid stress-relaxation time is $\tau=1.65 {\rm s}$, yielding the Deborah number ${\rm De}=U_0\tau/2R=0.05$. As reported in experiments \cite{gomez2016dynamics,narinder2018memory}, the memory effect resulting from polymer deformation is associated with fluid viscoelasticity, 
but the small ${\rm De}$ indicates the predominant viscous nature of their solutions. In contrast, our systems possess a very large Deborah number ${\rm De}=20944$ at ${\rm Re=0.04}$ and $k_BT/E_t=0.15$. However, it does not necessarily imply a strong elasticity resulting from the deformed polymers. 
In fact, the relaxation process of polymers is exceptionally slow, preventing squirmers from experiencing any noticeable elastic effect. In our work, the profound enhancements in the rotational motion are due to the mechanical and hydrodynamic interactions between polymers and swimmers during self-propulsion. Here, system heterogeneity plays a crucial role in determining the behavior, where polymers wrapping around the pusher generate torques to accelerate rotation, and backwards suction flows lead to polymer adsorption in the rear of a puller, inducing asymmetric flows that enhance rotation through the fluid-squirmer no-slip boundary condition.

On the other hand, a novel mechanism involving the spontaneous chiralization of polar active particles was proposed by Corato \cite{de2021spontaneous}. According to their work, a transient perturbation of the particle's polar vector breaks the symmetry of the solute distribution in the vicinity of the swimmer and thus induces a torque acting on the fluid. This torque is counterbalanced by an opposing torque on the particle. If the solute diffusion is slower than the advection generated by the self-propulsion, the resultant torque will be reinforced to accelerate the particle's rotation \cite{de2021spontaneous}. An extension of his work reveals that the interplay between active motion and fluid microstructure induces the spontaneous rotation of the Janus particle and significantly increases the rotational diffusion coefficient by orders of magnitude compared to its equilibrium value \cite{de2025enhanced}. The spontaneous chirality of a swimmer is characterized by solute concentration $\phi$ and the P\'eclet number associated with squirmer advection and polymer diffusion ${\rm Pe}_{p}=U_0R/D_t$, where $D_t$ is the translational diffusion coefficient of the solute. In our systems at temperature $k_BT/E=0.15$ and monomer concentration $R^3\phi=1.67$, $D_t=3.9\times10^{-5}, 4.5\times 10^{-6} a^2 \tau^{-1}$ in the finite and low Reynolds number regimes, respectively. The corresponding P\'eclet numbers are ${\rm Pe}_{p}=256,2222$. It is worth noting that even larger ${\rm Pe}_p$ can be obtained at $k_BT/E_t=0.0015$ due to the substantial decrease of solute diffusion. Compared with the phase diagram in \cite{de2021spontaneous}, our systems are in the polar regime, indicating that the spontaneous chiralization of a squirmer is not the primary mechanism
in determining the rotational behavior here.

Recently, a simulation work investigated the translational and rotational behavior of squirmers in flexible and semiflexible polymer solutions via multiparticle collision dynamics \cite{zottl2023dynamics}. Large reductions of swimming speeds for at most $40\%$ and $90\%$ in flexible and semi-flexible polymer solutions were reported, respectively. However, no evident enhancements of rotational diffusion were found. In their work, the Reynolds number ${\rm Re}=0.1$ is comparable to our ${\rm Re}= 0.04$ case. Meanwhile, the P\'eclect number ${\rm Pe}_t=3U_0\tau_r/(2R)\approx200$ was employed to characterize the transport behavior of the squirmer. Here the squirmer’s persistence time \cite{zottl2023dynamics} $\tau_r=4\pi R^3\eta/(k_B T)$ was used. In our case, the corresponding persistence times are $\tau_r=8.5 \times 10^{4}$ and $1.7 \times 10^6$ for $\eta=0.025$ and $0.5$ at temperature $T=10^{-4}$, resulting in the P\'eclect number ${\rm Pe}_t=142$ and $2833$, respectively. Therefore, the reduction of swimming speeds and differences in the squirmer’s rotational diffusion are not attributed to the dimensionless numbers employed in the two simulations. As a matter of fact, the translational kinetic energy of a squirmer passes to polymers and surrounding fluids via the repulsive interaction and the no-slip boundary condition, respectively. However, inertialess subgrid particles are used in our work to form a polymer chain. Therefore, the kinetic energy can only be distributed to the surrounding fluids either directly through the squirmer-fluid no-slip boundary condition or the interpolation mechanism applied to subgrid particles. As a consequence, at most $10\%$ reduction in the swimming speed is observed. Interestingly, no apparent enhancement in the rotational diffusion was observed in \cite{zottl2023dynamics}. This is probably attributed to the significant thermal fluctuations within their system, where a relatively high temperature $k_BT/E_t=0.43$ was applied. In contrast, a substantially low temperature $k_BT/E_t=0.0016$ is utilized in our work, resulting in pronounced rotational enhancements. It is important to note that excessively strong thermal fluctuations can obscure the flow inhomogeneities that accumulate due to the presence of polymers surrounding the squirmer. Consequently, no evident enhancement can be observed. It is noteworthy that squirmer-polymer attraction is not necessary for the occurrence of enhancement, which is still evident when the repulsion is utilized here.

The transport and rotational properties of microorganisms in viscoelastic environments are complex. Fascinating phenomena, such as accelerated swimming velocity and enhanced rotational motion, typically arise from the interplay of multiple factors. Our work highlights the crucial role of mechanical and hydrodynamic interactions between swimmers and polymers in shaping the behavior of squirmers in polymer solutions. In particular, the system heterogeneity induced by dispersed polymers plays an essential role. Large Deborah numbers obtained in our simulations reveal that the relaxation process of locally deformed polymers is too slow to exhibit any elastic effect. However, it is worth noting that further investigation is required to determine whether the entangled polymer solution as a whole possesses significant viscoelasticity.

\section{Acknowledgements}

The authors appreciate the support from the Swiss National Science Foundation program ``Computational modeling at CECAM. Challenges in the foundations and modeling of systems far away from equilibrium (200021\_175719)'' and the National Natural Science Foundation of China (No. 12304257). The authors gratefully acknowledge the computing time granted on the supercomputer Piz Daint at Centro Svizzero di Calcolo Scientifico CSCS.


\bibliographystyle{apsrev4-2}
\bibliography{squ_poly_LB}

\end{document}